\documentclass[sigconf,balance=false]{acmart}
\usepackage{popets}

\usepackage{amssymb}
\usepackage{algorithm}
\usepackage[noend]{algorithmic}
\usepackage{utfsym}
\usepackage{booktabs}       
\usepackage{amsfonts}       
\usepackage{nicefrac}       
\usepackage{microtype}      
\usepackage{xcolor}         
\usepackage{utfsym}
\usepackage{multirow}
\usepackage{makecell}       
\usepackage{amsmath}
\usepackage{pifont}
\usepackage{subfigure}
\usepackage{bbm}
\usepackage{listings}
\usepackage{paralist}
\usepackage{dsfont}
\usepackage{graphicx}
\usepackage{pgfplots}
\pgfplotsset{compat=1.18}
\usepackage{array}
\usepackage{mathtools}
\usepackage{amsthm}
\usepackage[pro]{fontawesome5}
\usepackage{enumitem}

\setcopyright{popets}
\copyrightyear{YYYY}

\acmYear{YYYY}
\acmVolume{YYYY}
\acmNumber{X}
\acmDOI{XXXXXXX.XXXXXXX}
\acmISBN{}
\acmConference{Proceedings on Privacy Enhancing Technologies}
\begin{document}

\title{SoK: Can Fully Homomorphic Encryption Support General AI Computation? A Functional and Cost Analysis}

\author{Jiaqi Xue}
\affiliation{
  \institution{University of Central Florida}
  \city{} 
  \state{} 
  \country{} 
}

\author{Xin Xin}
\affiliation{
  \institution{University of Central Florida}
  \city{}
  \country{}}

\author{Wei Zhang}
\affiliation{
  \institution{University of Central Florida}
  \city{}
  \country{}
}

\author{Mengxin Zheng}
\affiliation{%
 \institution{University of Central Florida}
 \city{}
 \state{}
 \country{}}

\author{Qianqian Song}
\affiliation{%
  \institution{University of Florida}
  \city{}
  \state{}
  \country{}}

\author{Minxuan Zhou}
\affiliation{
  \institution{Illinois Institute of Technology}
  \city{}
  \state{}
  \country{}}

\author{Yushun Dong}
\affiliation{%
  \institution{Florida State University}
  \city{}
  \country{}}

\author{Dongjie Wang}
\affiliation{
  \institution{The University of Kansas}
  \city{}
  \country{}}

\author{Xun Chen}
\affiliation{
  \institution{Samsung Research America}
  \city{}
  \country{}}

\author{Jiafeng Xie}
\affiliation{
  \institution{Villanova University}
  \city{}
  \country{}}

\author{Liqiang Wang}
\affiliation{%
 \institution{University of Central Florida}
 \city{}
 \state{}
 \country{}}

\author{David Mohaisen}
\affiliation{%
 \institution{University of Central Florida}
 \city{}
 \state{}
 \country{}}

\author{Hongyi Wu}
\affiliation{%
 \institution{University of Arizona}
 \city{}
 \state{}
 \country{}}

\author{Qian Lou\textsuperscript{$*$}}
\affiliation{%
 \institution{University of Central Florida}
 \city{}
 \state{}
 \country{}}

\renewcommand{\shortauthors}{Jiaqi Xue et al.}

\begin{abstract}

Artificial intelligence (AI) increasingly powers sensitive applications in domains such as healthcare and finance, relying on both \textit{linear operations} (e.g., matrix multiplications in large language models) and \textit{non-linear operations} (e.g., sorting in retrieval-augmented generation). Fully homomorphic encryption (FHE) has emerged as a promising tool for privacy-preserving computation, but it remains unclear whether existing methods can support the full spectrum of AI workloads that combine these operations.

In this SoK, we ask: \textit{Can FHE support general AI computation?} We provide both a functional analysis and a cost analysis. First, we categorize ten distinct FHE approaches and evaluate their ability to support general computation. We then identify three promising candidates and benchmark workloads that mix linear and non-linear operations across different bit lengths and SIMD parallelization settings. Finally, we evaluate five real-world, privacy-sensitive AI applications that instantiate these workloads. Our results quantify the costs of achieving general computation in FHE and offer practical guidance on selecting FHE methods that best fit specific AI application requirements. Our
codes are available at \href{https://github.com/UCF-ML-Research/FHE-AI-Generality}{https://github.com/UCF-ML-Research/FHE-AI-Generality}.
\end{abstract}


\maketitle

\section{Introduction}


The rapid advancement of AI has enabled its widespread adoption across privacy-sensitive domains, including healthcare~\cite{kim2015private, raisaro2018protecting, zhang2015foresee} and finance~\cite{armknecht2015guide, han2019logistic}. To protect privacy, FHE is a promising tool, as it enables direct computation on encrypted data without decryption, ensuring end-to-end confidentiality. However, AI workloads rely on both linear and non-linear operations, while existing FHE schemes are specialized rather than general: word-wise schemes (e.g., BGV~\cite{brakerski2014leveled, yudha2024boostcom}, BFV~\cite{brakerski2012fully, fan2012somewhat}, CKKS~\cite{cheon2017homomorphic}) are efficient for linear operations but struggle with non-linear ones, whereas bit-wise schemes (e.g., FHEW~\cite{ducas2015fhew}, TFHE~\cite{chillotti2020tfhe}) handle non-linear operations well but incur prohibitively high costs on linear ones. Thus, it remains unclear whether existing FHE can truly support the full spectrum of AI workloads that mix linear and non-linear operations in practice. For example, neural networks~\cite{gilad2016cryptonets,lou2021safenet} alternate linear matrix multiplications with non-linear activation functions, and graph algorithms such as Floyd–Warshall~\cite{floyd1962algorithm} combine linear aggregation with non-linear comparisons.

We refer to such mixed computations as general computation. The ability of an FHE method to support general computation in a non-interactive and efficient manner is what we call \textit{computational generality}. Although prior research has systematized FHE along dimensions such as performance, toolchain maturity, and developer accessibility~\cite{viand2021sok, gouert2023sok}, computational generality has received little attention—despite being essential for real AI applications. The open research question is twofold: \textit{first, the functionality of different FHE methods in supporting general computation, and second, the cost of executing it when such support is available.}



\begin{figure}[t!]
\centering
\includegraphics[width=0.65\linewidth]{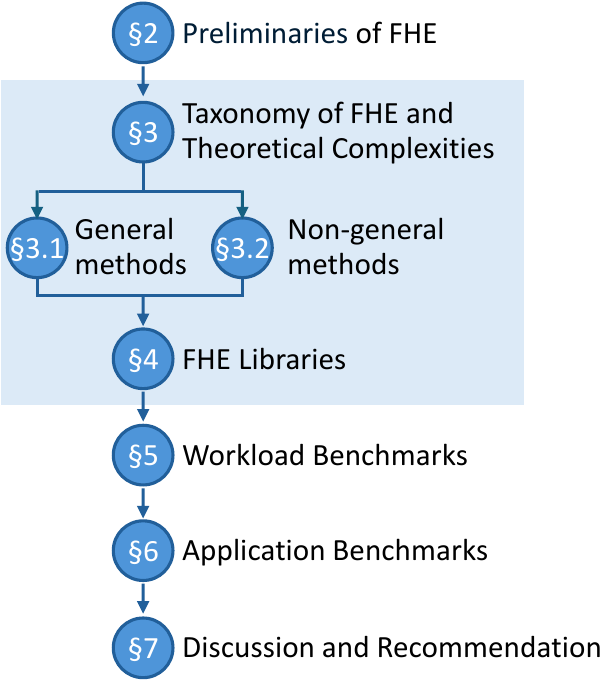}
\caption{Overview of Our FHE Generality Measurements.}
\label{f:over_SoK}
\end{figure}

\def\thefootnote{$*$}\footnotetext{Corresponding Author Email: qian.lou@ucf.edu.}


Given the diverse landscape of FHE methods that are specifically optimized for different applications and workloads, evaluating the FHE's computational generality is challenging. Many FHE methods are theoretically capable of general computation but remain impractical due to significant computational complexity.
For instance, directly using bit-wise FHE for general computation is impractical because of its extremely high computational complexity for linear computation~\cite{lu2021pegasus}, i.e., multiplying two 16-bit integers in TFHE can take up to 30 seconds. On the other hand, using word-wise FHE for non-linear functions with linear approximation appears to be a potential direction for handling general computation, such as polynomial approximation~\cite{cheon2020efficient, cheon2019numerical, lee2021minimax} for CKKS. However, these methods only work on a small interval, such as \([-1,1]\), inevitably introducing errors around 0. Such errors render approximation-based methods unsuitable for applications where even small errors are intolerable, such as genomics~\cite{kim2015private, raisaro2018protecting, zhang2015foresee} and finance~\cite{armknecht2015guide, han2019logistic}.
Another line of work for general computation involves scheme switching~\cite{lu2021pegasus, boura2020chimera} between word-wise and bit-wise FHE, allowing for the evaluation of the linear operations in word-wise and the non-linear operations in the bit-wise FHE without decryption. While this leverages the strengths of both types, switching cost is prohibitively high, especially for input with a large bit-width~\cite{cheon2022efficient}. Most importantly, additional bootstrapping might be needed after scheme switching~\cite{al2022openfhe}, which is one of the most computationally expensive operations in the word-wise FHE and is often avoided in practice~\cite{iliashenko2021faster}.

Additionally, while word-wise BGV/BFV can leverage \(\mathbb{F}_p\)-based polynomial interpolation~\cite{tan2020efficient, iliashenko2021faster, morimura2023accelerating} to enable exact non-linear operations, their efficacy in general computations remains unverified, because the inherent field disparity between \(\mathbb{F}_p\) (non-linear) and \(\mathbb{Z}_{p^d}\) (linear) fundamentally limits seamless integration of mixed operations.
To enable them to support general computation in a non-interactive manner, Zhang et al.~\cite{zhang2023hebridge} proposed an encoding switching method, HEBridge.
Recently, several functional bootstrapping methods~\cite{alexandru2024general, lee2024functional, liu2023amortized} have been proposed to allow arbitrary function computation during the bootstrapping process. Functional bootstrapping refreshes ciphertexts and applies a chosen function simultaneously, effectively removing noise while completing a specific Look-Up Table (LUT) evaluation. Specifically, Lee et al.~\cite{lee2024functional} proposed a functional bootstrapping technique that takes an RLWE ciphertext, i.e., CKKS or BFV, as input and outputs the refreshed BFV ciphertext, building upon BFV-style bootstrapping. However, such BFV-style bootstrapping is far from efficient; for instance, it can take about 3 minutes for a single evaluation due to the intrinsically inefficient slot utilization of BFV-style bootstrapping~\cite{alexandru2024general}. To address this issue, Alexandru et al.~\cite{alexandru2024general} presented a general functional bootstrapping technique based on CKKS-style bootstrapping, which has the best throughput among all FHE methods, to improve amortized performance. However, this method suffers from high computational complexity for large input spaces. Their experimental results are limited to 12-bit LUTs because the scaling factor nears the limit for 64-bit modular operations, and the complexity of polynomial evaluation increases significantly, i.e., approximately 1 minute per evaluation for 9-bit LUTs, but about 10 minutes for 12-bit LUTs.



To evaluate the computational generality of FHE, we proceed in three stages, as shown in Figure~\ref{f:over_SoK}. First, we survey and analyze current FHE methods to determine whether they have the functionality to support general computation. For FHE methods that support general computation, we analyze their theoretical complexities and identify the three most efficient candidates. Next, we conduct micro-benchmarks on three representative general workloads involving mixed linear and non-linear operations: (1) linear followed by non-linear, (2) non-linear followed by linear, and (3) a sequence of linear, non-linear, and linear operations. Finally, we extend our evaluation to five real-world, privacy-sensitive applications that inherently require FHE computational generality, including Floyd–Warshall graph analysis, decision tree inference, sorting, database aggregation, and neural network inference. These applications span domains such as finance and healthcare, where protecting sensitive information during complex computations is critical. By integrating theoretical analysis with empirical runtime measurements, we deliver a comprehensive, application-driven benchmark of general FHE solutions.

\section{Preliminaries}
\subsection{Fully Homomorphic Encryption}
Fully Homomorphic Encryption (FHE) is an encryption technique that enables computations to be carried out directly on encrypted data, ensuring confidentiality throughout processing. Formally, if \(y = f(x)\) represents an arithmetic function on plaintext \(x\), there exists a function \(g(\cdot)\) operating in the encrypted domain such that \(y = f(x) \Leftrightarrow y = \mathrm{Dec}(g(\mathrm{Enc}(x)))\). \(\mathrm{Enc}(\cdot)\) and \(\mathrm{Dec}(\cdot)\) denote encryption and decryption, respectively, while \(g(\cdot)\) mirrors \(f(\cdot)\) in the ciphertext space.
FHE methods introduce noise into ciphertexts to ensure security, typically following the Learning With Errors (LWE) problem. The noise prevents cryptanalysis but must remain below a threshold to ensure successful decryption. The security of modern FHE methods relies on the hardness of LWE and its ring-based variant, Ring LWE (RLWE). In RLWE, the goal is to find the secret \(s \in R_p\) in the equation \(b = a \cdot s + e\), where \(b, a \in R_p\) are known, and \(e\) is an error sampled from a distribution over \(R_p\). The ring \(R_p\) is defined as \(R_p = \mathbb{F}_p[x] / \langle x^n + 1 \rangle\), with \(n\) a power of 2 and \(p \equiv 1 \pmod{2n}\).
In practice, ciphertexts are encrypted as tuples of polynomials in this ring, taken modulo an irreducible cyclotomic polynomial whose roots are the \(N^\text{th}\) primitive roots of unity. The cyclotomic order (degree of the ciphertext polynomials) is typically between \(2^{10}\) and \(2^{15}\), and the coefficient modulus \(q\) is often a product of primes, reaching several hundred bits. 


Most current FHE methods can be categorized according to their capability to support different types of operations. Word-wise FHE methods (such as BFV~\cite{fan2012somewhat}, BGV~\cite{brakerski2014leveled}, and CKKS~\cite{cheon2017homomorphic}) excel at linear operations such as matrix multiplication, while bit-wise FHE methods (such as TFHE~\cite{chillotti2020tfhe} and FHEW~\cite{ducas2015fhew}) are better suited for non-linear operations such as comparison.

\noindent\textbf{Bit-wise FHE methods.} The first category includes bit-wise FHE methods, such as TFHE~\cite{chillotti2020tfhe} and FHEW~\cite{ducas2015fhew}, which encrypt individual bits into separate ciphertexts, allowing for a wide range of encrypted bitwise manipulations and logic gate evaluations. These schemes support the construction of Boolean circuits composed of encrypted logic gates, facilitating efficient operations directly on encrypted data. TFHE, in particular, is known for its fast bootstrapping, which is essential for evaluating homomorphic logic gates. Bit-wise FHE methods enable developers to leverage decades of digital circuit design research, enabling many applications such as privacy-preserving machine learning and relevant TFHE accelerators~\cite{lou2019she, lou2019glyph, jiang2022matcha, zheng2023priml, zheng2024ofhe, kumar2025tfhe, han2022coxhe, zheng2022cryptolight, zhu2025dahe, lou2019autoq, deng2024trinity}. However, they face challenges with addition and multiplication circuits~\cite{lou2019she, lu2021pegasus}, especially when dealing with large circuit depths and fan-in bits. For example, TFHE~\cite{chillotti2020tfhe} takes about 30 seconds to multiply two encrypted 16-bit integers. Also, the expansion ratio of bit-wise FHEs is usually several orders of magnitude larger than that of word-wise FHEs, leading to higher communication costs.

\noindent\textbf{Word-wise FHE methods.} Word-wise FHEs can be classified into integer schemes~\cite{brakerski2014leveled, brakerski2012fully, fan2012somewhat} and floating-point schemes~\cite{cheon2017homomorphic}. The two most important FHE schemes that encrypt integers modulo a user-determined modulus \(p\) are BGV~\cite{brakerski2014leveled} and BFV~\cite{fan2012somewhat}. FHE addition and multiplication correspond to modular addition and multiplication over the plaintext values. It is also possible to emulate Boolean circuits in integer schemes by setting \(p=2\), which causes addition to behave as an \texttt{XOR} gate and multiplication to work as an \texttt{AND} gate. Unlike bit-wise FHE schemes, both BGV and BFV have considerably slower bootstrapping; depending on parameter choices (such as the ring dimension), a single bootstrapping can take anywhere from several seconds to several hours~\cite{geelen2023bootstrapping}. Thus, most implementations of BGV and BFV do not include bootstrapping and are used exclusively in LHE mode~\cite{gouert2023sok}.

For the floating-point schemes, they use a plaintext type of floating-point numbers, and the most popular scheme in this category is CKKS~\cite{cheon2017homomorphic}. It is desirable for certain classes of algorithms, such as machine learning~\cite{xue2024cryptotrain,zhang2025cipherprune, zheng2023primer, zhao2025design, zhang2024heprune}. In practice, the core operations of floating-point schemes parallel those of integer schemes, with only a few differences. Similarly, the floating-point schemes have slow bootstrapping procedures and are predominantly used in LHE mode. The key difference is the need to keep track of the \textit{scale} factor that is multiplied by plaintext values during encoding, which determines the bit-precision associated with each ciphertext. The scale doubles when two ciphertexts are multiplied, which may result in overflow as the scale becomes exponentially larger. Thus, \textit{rescaling} must be invoked to preserve the original scale after multiplications, which is similar to modulus switching and serves a similar role in reducing ciphertext noise.

The key advantage of word-wise FHE over bit-wise methods is its support for batching. Single Instruction Multiple Data (SIMD) enables encrypting a vector of plaintexts into a single ciphertext~\cite{smart2014fully, cheon2017homomorphic}. Each plaintext occupies a \textit{slot}, with the number of slots determined by encryption parameters. Batched ciphertexts support slot-wise addition and multiplication. A key limitation is slot dependency---if operations require interaction across slots (e.g., summing all slots), slot-wise rotations are needed, incurring significant memory and runtime overhead.

Despite their efficiency in linear operations, word-wise FHE methods struggle with non-linear functions like comparison~\cite{lou2020falcon, lou2020autoprivacy}, prompting efforts to support such functions within word-wise schemes~\cite{iliashenko2021faster, tan2020efficient}. 
CKKS supports approximate computation on floating-point numbers, and the most prevailing method to perform the non-linear functions is through polynomial approximation~\cite{cheon2020efficient, cheon2019numerical, lee2021minimax}. These methods use a composition of minimax approximation polynomials~\cite{lee2021minimax} to approximate the sign function with errors. These methods primarily work on only small intervals, such as \([-1,-\epsilon]\cup[\epsilon, 1]\) with errors. The smaller the errors are, the higher the degree of the polynomial required. Thus, more multiplication and multiplicative depth are needed. Most importantly, the approximation can never be accurate around \(0\). Such errors limit the application of approximation-based non-linear methods where accuracy is important, such as genomics~\cite{kim2015private, raisaro2018protecting, zhang2015foresee} and finance~\cite{armknecht2015guide, han2019logistic}. 
On the other hand, since computations are exact in BFV and BGV, the non-linear functions can be implemented in BFV and BGV without approximation error. There has been a series of works~\cite{tan2020efficient, iliashenko2021faster, morimura2023accelerating} exploring the efficient comparison in BFV and BGV. On a high level, they compute the non-linear functions by evaluating an interpolation polynomial over the base field. However, the degree of the interpolation polynomial grows exponentially with the bit-width of the input~\cite{tan2020efficient} or relies on a special plaintext space~\cite{iliashenko2021faster} that does not support seamless computation of the linear and non-linear operations. 

\subsection{Bootstrapping Overview}
\label{sec:bootstrapping}
Bootstrapping is a fundamental operation in HE that refreshes noisy ciphertexts to enable unbounded computation. Each homomorphic operation introduces noise into ciphertexts, and once noise exceeds the decryption threshold, correctness fails. Bootstrapping homomorphically evaluates the decryption function on a noisy ciphertext, producing a refreshed ciphertext with reduced noise: \(\textbf{ct}'=\text{Enc}_\text{sk}'(\text{Dec}_{\text{sk}}(\textbf{ct}))=\text{Enc}_\text{sk}'(m)\). The three computationally general FHE methods employ fundamentally different bootstrapping strategies with dramatically different costs. Bit-wise TFHE uses gate bootstrapping (around \(10-20\) ms per gate), where each gate evaluation is immediately followed by noise refresh, enabling constant multiplicative depth but requiring separate bootstrapping for each bit. Word-wise methods (Scheme Switching and Encoding Switching) rely on BGV/BFV bootstrapping, which is orders of magnitude slower (3-30 seconds per operation) but can amortize costs across thousands of SIMD slots, achieving ~1-7 ms per slot when batching is possible. Due to these prohibitive word-wise bootstrapping costs, most implementations operate in leveled mode with sufficiently large parameters to avoid bootstrapping entirely.

\subsection{Generality Required Privacy-sensitive Applications}

Over the past decade, FHE has evolved from a purely theoretical construct into a practical solution for today’s data-centric challenges~\cite{gouert2023sok}. In finance, Intesa Sanpaolo's collaboration with IBM leverages FHE secure asset transactions~\cite{finance_exp}, allowing for both confidential database querying and the execution of complex machine learning algorithms on encrypted data. Similarly, in the medical domain, Roche’s partnership with ETH investigates the use of FHE for secure graph analyses of genomic information~\cite{lou2024homomorphic,hipaa}. As industries increasingly adopt advanced analytics that require both linear and non-linear operations, FHE’s generality to handle complex computations on encrypted data makes it an indispensable tool for ensuring data confidentiality and regulatory compliance.



In Section~\ref{sec:analysis_theoretical}, we analyze the computational generality of various FHE methods from a theoretical perspective. Section~\ref{sec:workloads} presents experimental evaluations assessing their performance across diverse workloads. These workloads include sequences of operations such as non-linear applied after linear operations, mirroring the activation functions in neural networks; as well as linear following non-linear operations, which form the basic unit of basic \(\text{Max}/\text{Min}(A,B)\) operations, expressed as \( \text{Compare}(A, B) \times A + (1 - \text{Compare}(A, B)) \times B \).
Finally, in Section~\ref{sec:applications}, we evaluate privacy-sensitive AI-related applications built on these workloads, demonstrating the practical generality of FHE methods. The applications under study include graph algorithms, decision trees, database retrieval, neural-network inference, and sorting, each representing a critical function in real-world research and commercial settings.





\section{FHE Generality Taxonomy and Complexity Analysis}\label{sec:analysis_theoretical}

\begin{table*}[ht!]
\caption{\textcolor{black}{Functionality and cost comparison of representative FHE methods categorized by functional capability and theoretical complexity under $b$-bit precision and $2^{\lambda}$ security level. The last three lines represent the methods achieve full computational generality across linear and non-linear operations. $N$ is the ring dimension; for polynomial approximation of CKKS, $n$ denotes the polynomial degree.}}
\label{properties-comparison}
\centering
\footnotesize
\setlength{\tabcolsep}{2.5pt}
\begin{tabular}{lccccccccccc}\toprule
\multirow{2}{*}{Methods} & \multicolumn{5}{c}{Non-linear Operation (i.e., Comparison)} & \multicolumn{4}{c}{Linear Operation (i.e., Multiplication)} & \multirow{2}{*}{Generality}\\\cmidrule(lr){2-6}\cmidrule(lr){7-10}
& Depth & Complexity & SIMD & Exact & Efficiency & Depth & Complexity & SIMD & Efficiency \\\midrule
Word-wise BGV/BFV~\cite{brakerski2014leveled, brakerski2012fully} & \textbf{-} & \textbf{-} & \textbf{-} & \textbf{-} & \textbf{-} & $1$ & $O(\lambda^2)$ & \textcolor{black}{\ding{51}} & \textcolor{black}{\ding{51}} & \textcolor{black}{\ding{55}} \\
Word-wise CKKS~\cite{cheon2017homomorphic} & \textbf{-} & \textbf{-} & \textbf{-} & \textbf{-} & \textbf{-} & $1$ &$O(N\log N)$ & \textcolor{black}{\ding{51}} & \textcolor{black}{\ding{51}} & \textcolor{black}{\ding{55}} \\

Poly. Approx. (CKKS)~\cite{cheon2020efficient, lee2021minimax} & $O(\log_2n)$ & $O(\sqrt{n})$  & \textcolor{black}{\ding{51}} & \textcolor{black}{\ding{55}} & \textcolor{black}{\ding{51}} & $1$ &$O(N\log N)$& \textcolor{black}{\ding{51}} & \textcolor{black}{\ding{51}} & \textcolor{black}{\ding{55}} \\
Poly. Interp. (BGV/BFV)~\cite{narumanchi2017performance} & $O(b)$ & $O(\sqrt{2^b})$ & \textcolor{black}{\ding{51}} & \textcolor{black}{\ding{51}} & \textcolor{black}{\ding{55}} & $1$ & $O(\lambda^2)$ & \textcolor{black}{\ding{51}} & \textcolor{black}{\ding{51}} & \textcolor{black}{\ding{55}} \\
Decomp. (BGV/BFV)~\cite{iliashenko2021faster} & $O(\log_2b)$ & $O(b\log_2b)$ & \textcolor{black}{\ding{51}} & \textcolor{black}{\ding{51}} & \textcolor{black}{\ding{51}} & \textbf{-} & \textbf{-} & \textbf{-} & \textbf{-} & \textcolor{black}{\ding{55}} \\
XCMP (BGV/BFV)~\cite{lu2018non} & $O(1)$ & $O(1)$ & \textcolor{black}{\ding{51}} & \textcolor{black}{\ding{51}} & \textcolor{black}{\ding{51}} & \textbf{-} & \textbf{-} & \textbf{-} & \textbf{-} & \textcolor{black}{\ding{55}} \\
General Functional Bootstrapping~\cite{alexandru2024general} & $O(b)$ & $O(\sqrt{2^b}+b)$ & \textcolor{black}{\ding{51}} & \textcolor{black}{\ding{51}} & \textcolor{black}{\ding{55}} & $1$ &$O(N\log N)$ & \textcolor{black}{\ding{51}} & \textcolor{black}{\ding{51}} & \textcolor{black}{\ding{55}} \\\midrule

Bit-wise TFHE~\cite{chillotti2020tfhe} & \(O(b)\) & \(O(b)\) & \textcolor{black}{\ding{55}} & \textcolor{black}{\ding{51}} & \textcolor{black}{\ding{51}} & $O(b)$ & $O(b^2p) \textit{CMUX}+ O(bp) \textit{KS} $ & \textcolor{black}{\ding{55}} & \textcolor{black}{\ding{55}} & \textcolor{black}{\ding{51}} \\

Scheme Switching~\cite{lu2021pegasus, boura2020chimera} & \textbf{-} & $\Omega(2^b)$ & \textcolor{black}{\ding{55}} & \textcolor{black}{\ding{51}} & \textcolor{black}{\ding{55}} & $1$ & $O(\lambda^2)$ & \textcolor{black}{\ding{51}}& \textcolor{black}{\ding{51}} & \textcolor{black}{\ding{51}} \\
Encoding Switching~\cite{zhang2023hebridge} & $O(\log_2d+d\log_2p)$ & $O(d^2\sqrt{p})$ & \textcolor{black}{\ding{51}} & \textcolor{black}{\ding{51}} & \textcolor{black}{\ding{51}} & $O(1)$ & $O(\lambda^2)$ & \textcolor{black}{\ding{51}} & \textcolor{black}{\ding{51}} & \textcolor{black}{\ding{51}} \\
\bottomrule
\end{tabular}
\label{t:overview_u}
\end{table*}

General computation is a crucial requirement of real-world privacy-sensitive applications, yet little research has investigated how existing FHE methods support this functionality. Specifically, an AI-friendly FHE method must satisfy two conditions:

\noindent\textbf{Mixed Linear and Non-Linear Operations:} An FHE method with computational generality must support both linear and non-linear operations on ciphertexts and allow them to be composed in arbitrary sequences. This capability enables complex computations that alternate between operation types, such as machine learning algorithms that apply linear operations (e.g., matrix multiplications) followed by non-linear activation functions (e.g., ReLU). By seamlessly mixing linear and non-linear operations, such FHE methods can accommodate a wide range of applications.

\noindent\textbf{Non-Interactive Computation:} An FHE method with computational generality should support these mixed operations in a non-interactive manner, allowing computations on encrypted data without further communication between the data owner and the computing party to decrypt and re-encrypt. Non-interactivity is essential for scalability and efficiency, especially in settings where minimizing communication overhead is critical.

{\color{black}
Among existing FHE methods, only a few can simultaneously satisfy both of above conditions that can support both linear and non-linear computation in a non-interactive manner, what we term full computational generality. This section primarily focuses on these three computationally general FHE methods: bit-wise TFHE, Scheme Switching, and Encoding Switching. We analyze their theoretical complexities and evaluate their suitability for general AI computation in Section~\ref{sec: UFHE}.

Other FHE methods, such as native word-wise BGV/BFV/CKKS, polynomial approximation, polynomial interpolation, and functional bootstrapping, provide only partial or approximate forms of computational generality. We discuss these method in Section~\ref{sec: un-UFHE} for conceptual completeness and to contextualize our focus. Table~\ref{t:overview_u} summarizes the generality functionality and computational complexities of representative FHE methods, highlighting the distinction between general and un-general general approaches.
}


\subsection{Analysis of General FHE methods}
\label{sec: UFHE}

This section analyzes three FHE methods that achieve computational generality, supporting both linear and non-linear operations in a non-interactive manner. The first is bit-wise FHE\cite{chillotti2020tfhe, ducas2015fhew}, which inherently enables both operation types through arithmetic and boolean circuits. The second is scheme switching\cite{lu2021pegasus, boura2020chimera}, which combines bit-wise and word-wise FHE to leverage the strengths of each. The third is encoding switching (i.e., HEBridge~\cite{zhang2023hebridge}), which integrates linear and non-linear operations within word-wise BGV/BFV by bridging distinct encoding spaces through homomorphic reduction and lifting functions.

\noindent\textbf{Bit-wise FHE.} TFHE~\cite{chillotti2020tfhe} is a bit-wise FHE method that supports gate-level operations on encrypted data, enabling fast non-linear operations due to its efficient bit-wise processing. For a ciphertext with \(b\)-bit precision, TFHE can perform non-linear functions with a depth of \(O(b)\) and complexity of \(O(\sqrt{2^b})\). However, TFHE's linear operations are comparatively slower due to its bit-wise design. Multiplication on ciphertexts requires a depth of \(O(b^2)\), along with \(O(b^2p)\) CMUX operations and \(O(bp)\) key-switching. Additionally, TFHE lacks support for batch processing, i.e., SIMD, making linear computations less efficient than in word-wise FHEs, which can parallelize linear operations.

\noindent\textbf{Scheme Switching.} Scheme switching~\cite{boura2020chimera, lu2021pegasus} between bit-wise and word-wise ciphertexts enables general computation by leveraging the strengths of both schemes: linear functions are evaluated in word-wise FHE, and non-linear functions in bit-wise FHE.
However, the complexity of scheme switching grows at least exponentially with input bit-width~\cite{ren2022heda, bian2023he3db}, i.e., \(\Omega(2^b)\). For example, evaluating 6-bit inputs takes 43.8 seconds, while 8-bit inputs require 162.7 seconds. Consequently, scheme switching becomes impractical for larger inputs.

\noindent\textbf{Encoding Switching.} Prior studies~\cite{iliashenko2021faster, tan2020efficient} have shown that word-wise BGV/BFV support efficient and precise non-linear operations via polynomial interpolation. However, these methods cannot be seamlessly incorporated with linear operations because the plaintext space for linear operations in FV over \(\mathbb{Z}_{p^d}\) is different that used for non-linear operations in base-encoded FV (beFV) over \(\mathbb{F}_p\). To enable continuous evaluation of both operation types, Zhang et al.~\cite{zhang2023hebridge} introduced conversion techniques that employ a reduction function, which homomorphically converts the plaintext space from FV to beFV, allowing non-linear operations to follow linear operations, and a lifting function that restores the result from beFV back to the original FV space for subsequent linear operations. In terms of complexity, given that the FV plaintext modulus is \(p^d\), the reduction function requires \(O(d^2\sqrt{p})\) multiplications and \(O(d\log_2 p)\) multiplicative depth, while the lifting function requires \(O(\sqrt{dp})\) multiplications and a depth of \(O(\log_2 d+\log_2 p)\).

\subsection{Analysis of un-General FHE methods}
\label{sec: un-UFHE}
\textcolor{black}{While the three methods analyzed in Section~\ref{sec: UFHE} provide full computational generality, several other FHE approaches offer partial or approximate support for mixed linear and non-linear operations. We provide a brief overview of these methods here for conceptual completeness, emphasizing why they fall short of full generality.}

\noindent\textbf{Word-wise FHE Schemes.} Word-wise FHEs, including BGV, BFV, and CKKS, na\"{\i}vely support efficient linear operations with SIMD. However, computing non-linear functions in SIMD-enabled word-wise FHE is non-trivial. As a result, many efforts have been made to enable non-linear functions within word-wise FHE to leverage SIMD. For the CKKS scheme, the most common approach is polynomial approximation~\cite{cheon2020efficient, cheon2019numerical, lee2021minimax}. On the other hand, for integer-based BGV and BFV, Narumanchi et al.~\cite{narumanchi2017performance} proposed evaluating non-linear functions via polynomial interpolation over the base field.

{\color{black}
\noindent\textbf{CKKS with Polynomial Approximation.}
CKKS can approximate non-linear functions (e.g., sign, ReLU, sigmoid) using compositions of minimax approximation polynomials over small intervals~\cite{lee2023precise, lee2022privacy} such as \([-1, 1]\). Given an $n$-degree approximation polynomial, the complexity is \(O(\sqrt{n})\) and the multiplicative depth is \(\log_2n\). Higher polynomial degrees reduce approximation error but increase computational cost and multiplicative depth.

\noindent\textit{Limitation for General Computation:} Polynomial approximation introduces inherent errors, especially near zero, making it unsuitable for applications requiring exact computation, such as genomics and finance. While it provides a form of computational generality, it is approximate rather than exact, limiting its applicability in error-sensitive domains. More details are in the Appendix~\ref{app:CKKS}.

 



\noindent\textbf{BGB/BFV with Polynomial Interpolation.} 
In contrast to CKKS, computations in BGV/BFV are exact. Non-linear functions can be implemented via polynomial interpolation over the base field \(\mathbb{F}_p\). For example, comparison operations can be expressed using Lagrange interpolation polynomials evaluated homomorphically on encrypted differences~\cite{narumanchi2017performance}.

\noindent\textit{Limitation for General Computation:} While polynomial interpolation supports exact non-linear operations, the degree of the interpolation polynomial grows exponentially with input bit-width when performed over \(\mathbb{Z}_{p^r}\). To maintain efficiency, operations are typically confined to small primes \(p\leq257\), making this approach impractical for large inputs. Additionally, the plaintext space for linear operations (\(\mathbb{Z}_{p^d}\)) differs from that used for non-linear operations \(\mathbb{F}_p\), preventing seamless composition without encoding conversion (as in Encoding Switching~\cite{zhang2023hebridge}). More details are in the Appendix~\ref{app:Interpolation}.

\noindent\textbf{Polynomial Interpolation with Special Encoding.}
To enhance the scalability of polynomial interpolation, special encoding methods~\cite{iliashenko2021faster, tan2020efficient} decompose large integers into vectors of base-\(p\) digits. Non-linear operations (e.g., comparisons) are then performed digit-wise, reducing depth from \(\log_2{p^r}\) to \(\log_2\log_p2^b+\log_2(p-1)+4\).

\noindent\textit{Limitation for General Computation:} While these methods significantly improve the efficiency of non-linear operations, the specialized ciphertext format (vector encoding) does not support standard linear operations. Thus, they cannot seamlessly compose linear and non-linear operations without format conversion, disqualifying them as fully general FHE methods. More details are in the Appendix~\ref{app:SE}.

\noindent\textbf{Exponential Encoding (XCMP).} 
XCMP~\cite{lu2018non} uses exponential encoding to perform private comparisons with constant multiplicative depth \(O(1)\) by encoding values as polynomial degrees. This enables highly efficient comparisons but is limited to small input domains (typically < $16$ bits) due to the polynomial degree constraint.

\noindent\textit{Limitation for General Computation:} XCMP's specialized encoding precludes direct composition with linear operations, and extensions to larger domains incur prohibitive multiplicative depth. More details are in the Appendix~\ref{app:XCMP}.

\noindent\textbf{General Functional Bootstrapping.}
Functional bootstrapping~\cite{alexandru2024general, lee2024functional, liu2023amortized} extends traditional bootstrapping by enabling the evaluation of arbitrary functions (via look-up tables) during noise refresh. Methods such as BFV-style~\cite{lee2024functional} and CKKS-style~\cite{alexandru2024general} functional bootstrapping approximate target functions using interpolation polynomials with complexity roughly proportional to \(\sqrt{p}\) for first-order interpolation.

\noindent\textit{Limitation for General Computation:} While theoretically powerful, functional bootstrapping remains practical only for small input spaces (\(\leq 12\) bits) due to polynomial degree growth and scaling factor constraints. Moreover, extending to larger inputs requires digit-wise decomposition (similar to special encoding methods), which again precludes seamless linear-nonlinear composition. More details are in the Appendix~\ref{app:GB}.

}

\section{Relationship to Prior SoK work}
\textcolor{black}{Gouert et al.~\cite{gouert2023sok} presented a comprehensive SoK on FHE libraries, providing standardized benchmarks across major implementations including HElib, SEAL, Lattigo, OpenFHE, TFHE, and Concrete. Their work focuses on library-level performance evaluation—measuring runtime, memory consumption, ciphertext expansion, and parameter selection workflows across standard primitive operations (encryption, addition, multiplication, rotation). This library-focused SoK enables practitioners to compare implementation quality, identify performance bottlenecks, and select appropriate libraries for deployment. Additionally, Viand et al.~\cite{viand2021sok} primarily surveys FHE compilers, emphasizing performance, toolchain maturity, and developer accessibility. In contrast, our SoK addresses a fundamentally different research question: "Can existing FHE methods support general AI computation?" Rather than benchmarking library implementations, we analyze the functional generality and computational scalability of underlying FHE methods when applied to AI workloads requiring both linear and non-linear operations. Table~\ref{tab:sok-comparison} summarizes the key distinctions between these three complementary systematization efforts.
\begin{table}[h]
\centering
\scriptsize
\setlength{\tabcolsep}{2pt}
\caption{\textcolor{black}{Comparison of FHE SoK contributions.}}
\label{tab:sok-comparison}
\begin{tabular}{llll}
\toprule
\textbf{SoK} & \textbf{Focus} & \textbf{Units Evaluated} & \textbf{Contribution} \\
\midrule
Viand et al. & Compiler usability & Compilers (EVA, CHET, & Compiler landscape \\
\cite{viand2021sok} & \& performance & Cingulata, etc.) & systematization \\
\midrule
Gouert et al. & Library & Libraries (HElib, SEAL, & Standardized library \\
\cite{gouert2023sok} & performance & Lattigo, TFHE, PALISADE) & benchmarking \\
\midrule
\textbf{Our} & \textbf{Computational} & \textbf{FHE Methods (TFHE, Scheme} & \textbf{Method selection} \\
\textbf{Work} & \textbf{generality} & \textbf{Switching, Encoding Switching)} & \textbf{for general AI} \\
\bottomrule
\end{tabular}
\end{table}
} 

\section{FHE Libraries Supporting General Computation}



Various open-source FHE libraries implement the aforementioned FHE methods and provide high-level APIs that allow users to select encryption parameters tailored to their applications. In this section, we list six widely used libraries and summarize their support for these FHE methods.

Table~\ref{t:libraries} summarizes FHE libraries alongside the encryption schemes they implement and the types of supported operations. In general, HELib, Lattigo, and SEAL offer word-wise FHEs by supporting BFV, BGV and CKKS. In contrast, TFHE and Zama’s TFHE‑rs are designed primarily for bit-wise FHEs, offering Boolean gate-level primitives that can be composed into more complex functions. Notably, OpenFHE and Zama extend this generality further by facilitating Scheme Switching or hybrid approaches that bridge bit-wise and word-wise computations. Additionally, methods Polynomial Interpolation~\cite{narumanchi2017performance} and Polynomial Interpolation with Special Encoding~\cite{iliashenko2021faster} are open-sourced based on HELib, while Encoding Switching~\cite{zhang2023hebridge} is open-sourced using HELib. Functional bootstrapping techniques~\cite{alexandru2024general, lee2024functional} are implemented on Lattigo and OpenFHE but have not yet been open-sourced.

\begin{table}[h!]
\caption{Compatibility of libraries with various FHE methods.}
\centering
\footnotesize
\setlength{\tabcolsep}{2pt}
\begin{tabular}{lcccccc}\toprule
Methods & HElib & Lattigo & SEAL & TFHE & OpenFHE & Zama \\\midrule
Word-wise BGV & \textcolor{black}{\ding{51}} & \textcolor{black}{\ding{51}} & \textcolor{black}{\ding{51}} & \textcolor{black}{\ding{55}} & \textcolor{black}{\ding{51}} & \textcolor{black}{\ding{55}} \\
Word-wise BFV & \textcolor{black}{\ding{55}} & \textcolor{black}{\ding{51}} & \textcolor{black}{\ding{51}} & \textcolor{black}{\ding{55}} & \textcolor{black}{\ding{51}} & \textcolor{black}{\ding{55}} \\
Word-wise CKKS & \textcolor{black}{\ding{51}} & \textcolor{black}{\ding{51}} & \textcolor{black}{\ding{51}} & \textcolor{black}{\ding{55}} & \textcolor{black}{\ding{51}} & \textcolor{black}{\ding{55}} \\
Bit-wise TFHE & \textcolor{black}{\ding{55}} & \textcolor{black}{\ding{55}} & \textcolor{black}{\ding{55}} & \textcolor{black}{\ding{51}} & \textcolor{black}{\ding{51}} & \textcolor{black}{\ding{51}}\\
Poly. Interp. & \textcolor{black}{\ding{51}} & \textcolor{black}{\ding{55}} & \textcolor{black}{\ding{55}} & \textcolor{black}{\ding{55}} & \textcolor{black}{\ding{55}} & \textcolor{black}{\ding{55}} \\
Poly. Approx. & \textcolor{black}{\ding{51}} & \textcolor{black}{\ding{51}} & \textcolor{black}{\ding{51}} & \textcolor{black}{\ding{55}} & \textcolor{black}{\ding{51}} & \textcolor{black}{\ding{55}} \\
Scheme Switching & \textcolor{black}{\ding{55}} & \textcolor{black}{\ding{51}} & \textcolor{black}{\ding{55}} & \textcolor{black}{\ding{55}} & \textcolor{black}{\ding{51}} & \textcolor{black}{\ding{55}} \\
Encoding Switching & \textcolor{black}{\ding{51}} & \textcolor{black}{\ding{55}} & \textcolor{black}{\ding{55}} & \textcolor{black}{\ding{55}} & \textcolor{black}{\ding{55}} & \textcolor{black}{\ding{55}} \\
General Functional Bootstrapping  & \textcolor{black}{\ding{55}} & \textcolor{black}{\ding{51}\rotatebox[origin=c]{-9.2}{\kern-0.7em\ding{55}}} & \textcolor{black}{\ding{55}} & \textcolor{black}{\ding{55}} & \textcolor{black}{\ding{51}\rotatebox[origin=c]{-9.2}{\kern-0.7em\ding{55}}} & \textcolor{black}{\ding{55}} \\
\bottomrule
\end{tabular}
\label{t:libraries}
\end{table}

\noindent\textbf{HELib.} The Homomorphic Encryption Library (HELib) was introduced in 2013 by IBM and supports the BGV scheme (with bootstrapping), as well as CKKS. Polynomial Interpolation~\cite{narumanchi2017performance, iliashenko2021faster} with BGV and Encoding Switching~\cite{zhang2023hebridge} are implemented on HELib and have been open-sourced. HELib is written in C++17 and uses the NTL mathematical library. 

\noindent\textbf{Lattigo.} The lattice-based multiparty HE library in Go was first developed by the Laboratory for Data Security (LDS) at EPFL and is currently maintained by Tune Insight. It supports BFV, BGV, and CKKS. Lattigo enables scheme switching to compute non-linear functions. Functional FV bootstrapping~\cite{lee2024functional} is built on the lattigo, but their implementation is not open-sourced yet.

\noindent\textbf{SEAL.} The Simple Encrypted Arithmetic Library (SEAL) was developed by Microsoft Research and was first released in 2015~\cite{mircroSEAL}. SEAL supports leveled BFV, BGV, and CKKS.

\noindent\textbf{TFHE.} The Fast Fully Homomorphic Encryption Library over the Torus (TFHE) was released in 2016 by Chillotti et al.~\cite{chillotti2020tfhe} and proposes the CGGI cryptosystem. The library exposes homomorphic Boolean gates such as AND and XOR, but does not build complex functional units (e.g., adders, multipliers, and comparators) and leaves that to the developer.

\noindent\textbf{OpenFHE.} OpenFHE~\cite{al2022openfhe} is developed by Duality, NJIT, MIT, and other organizations. It supports a wide range of FHE methods, including BGV, BFV, and CKKS with approximate bootstrapping, as well as DM/FHEW, CGGI/TFHE, and LMKCDEY for evaluating Boolean circuits. OpenFHE enables scheme switching between CKKS and between CKKS and FHEW/TFHE to evaluate non-smooth functions, e.g., comparison, using FHEW/TFHE functional bootstrapping. Recently, Alexandru et al.~\cite{alexandru2024general} leveraged OpenFHE to build general functional bootstrapping to enable the non-linear function for any RLWE ciphertexts, but it is not open-sourced yet.

\noindent\textbf{Zama.} Zama~\cite{Zama} is an open-source cryptography company developing state-of-the-art FHE solutions for blockchain and AI. Its products include TFHE-rs~\cite{TFHE-rs} for Boolean and small integer arithmetic, Concrete~\cite{Concrete} for compiling Python to FHE with LLVM, Concrete ML~\cite{ConcreteML} for encrypted machine learning, and fhEVM for confidential smart contracts in Solidity.


\section{General Computation Required Workloads}
\label{sec:workloads}

\subsection{Design Principle}
To comprehensively evaluate and compare different general FHE methods, we designed three workloads requiring generality, each representing basic computational units in privacy-sensitive AI applications, covering diverse scenarios:
\begin{itemize}[leftmargin=*, nosep, topsep=0pt, partopsep=0pt, parsep=5pt]
\item Workload-1: \( \text{Compare}\left(\text{Enc}(A) \times \text{Enc}(B), \text{Enc}(C)\right) \), a non-linear operation following a linear operation that serves as a basic unit for database queries. For instance, it can be used to determine if the product of two encrypted values exceeds a given threshold.
\item Workload-2: \( \text{Compare}\left(\text{Enc}(A), \text{Enc}(B)\right) \times \text{Enc}(C) \), a linear operation following a non-linear operation that forms a basic unit in decision tree algorithms. For example, this unit can evaluate a comparison between two encrypted feature values and then multiply the result by another encrypted value to determine the weight or selection criteria at a decision node.
\item Workload-3: \( \text{Compare}\left(\text{Enc}(A) \times \text{Enc}(B), \text{Enc}(C)\right) \times \text{Enc}(D)\), a composite sequence of linear, non-linear, and linear operations. It is a basic component of a neural network, where it can model the execution of a convolution layer, followed by a ReLU activation, and then another convolution layer.
\end{itemize}

\subsection{Experimental Setup}
\label{sec:workload_setup}


To evaluate the performance of general FHE methods, we conducted experiments on the above workloads and measured both the total execution time and the amortized time for SIMD-enabled word-wise Scheme switching and Encoding switching.


\noindent\textbf{System Setup.} The experiments are conducted on a server equipped with an AMD Ryzen Threadripper PRO 3955WX (2.2 GHz) and 125 GB of RAM. All tests are run in single-thread mode for a fair comparison. For Encoding Switching, we use the official open-sourced implementation from HEBridge~\cite{zhang2023hebridge} built on HELib. Scheme Switching is adopted from OpenPEGASUS~\cite{lu2021pegasus} built on Microsoft SEAL, and TFHE-rs~\cite{TFHE-rs} are used for the bit-wise TFHE method. 
All encryption parameters are configured to maintain a security level above 128 bits~\cite{lauter2022protecting, bossuat2024security}, following the "BKZ-beta" classical cost model from the LWE estimator~\cite{albrecht2015concrete} unless stated otherwise. 

\noindent\textbf{Parameter Setup.} \textcolor{black}{Parameter choices ensure operations remain within the multiplicative depth budget, avoiding word-wise bootstrapping.} For both BGV/BFV‐based methods, i.e., Encoding Switching and Scheme Switching, we set the secret distribution to be a Hamming weight distribution over the set of ternary polynomials with coefficients in \(\{0,1,-1\}\), such that each secret has exactly \(h=64\) nonzero entries.

For Encoding Switching, which is built on HELib, the plaintext modulus in the FV space is given by \(p^r\). We choose the parameters \(p\) and \(r\) based on the input bit-width. Specifically, we use the pairs \(\{(4,4), (5,4), (7,5), (17,4)\}\) as \((p,r)\) for input bit-widths of \(6\), \(8\), \(12\), and \(16\), respectively. The multiplicative depth is determined by the ciphertext capacity, defined as \(\log_2\frac{q}{\eta}\), where \(q\) is the ciphertext modulus and \(\eta\) is the current noise bound. Accordingly, we set \(\log_2 q\) to \(\{256, 320, 488, 648\}\) for \(6\), \(8\), \(12\), and \(16\) bits of input. After choosing \(p\), \(r\), and \(q\), we select an appropriate degree for the polynomial modulus to ensure that the security level satisfies \(\lambda>128\). In this context, the cyclotomic order of the polynomial ring \(m\) is chosen so that the ring degree \(n\) and the order of the base prime, \(d=\texttt{Ord}(p)\), meet the following values: \(\{(m,n)=(13201,12852), (16151,15600), (25301,25300), (31621,31212)\}\), corresponding to \(6\), \(8\), \(12\), and \(16\) bits of input, respectively.

For Scheme Switching, which is implemented using Microsoft SEAL, we set the parameters in a way that mimics the above setup to ensure a fair comparison. Specifically, we explicitly configure the plaintext modulus \(p^r\) using the identical \((p,r)\) pairs, and we select the ciphertext modulus \(q\) with the same \(\log_2 q\). Then, we choose a polynomial modulus degree \(n\) (which in SEAL must be a power of two) such that the overall security level satisfies \(\lambda>128\). Although Microsoft SEAL does not allow explicit setting of the cyclotomic order \(m\) and the order of \(p\) (denoted as \(d\)), these are implicitly determined by our choices of \(n\) and \(p\) (with \(m\) being typically interpreted as \(2n\) for the cyclotomic polynomial \(x^n+1\), and \(d=\texttt{Ord}(p)\) accordingly).

For the bit-wise TFHE method, TFHE-rs provides \texttt{FheUint6}, \texttt{FheUint8}, \texttt{FheUint12}, \texttt{FheUint16}, \texttt{FheUint24}, and \texttt{FheUint32} to represent differnt input bit precision.

Our benchmarks focus exclusively on server-side FHE operations, including all evaluation tasks except key generation and encryption/decryption, as these are one-time user costs. No custom parallelization is applied to HElib, OpenFHE, or TFHE-rs, as each defaults to single-core execution.

\begin{figure*}[t!]
    \centering
        \includegraphics[width=\linewidth]{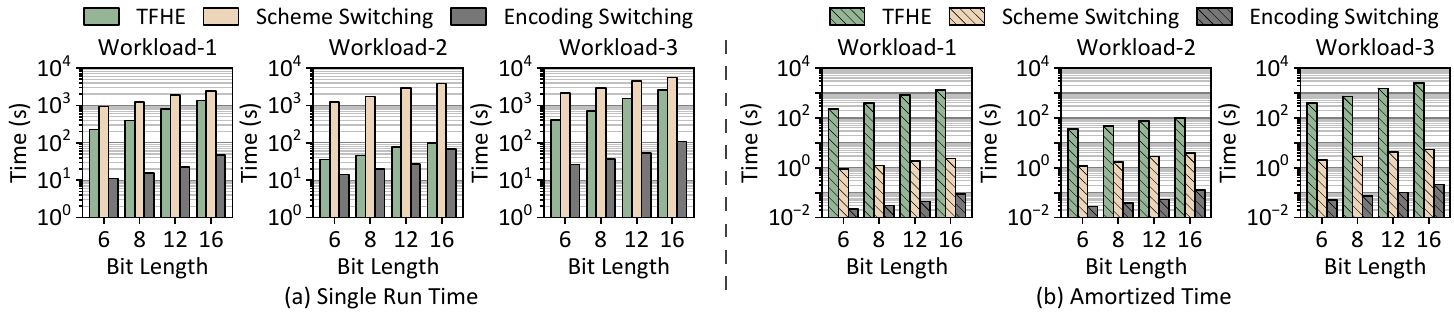}
        \caption{Comparisons on three workloads.}
        \label{f:comp_workloads}
\end{figure*}

\subsection{Experimental Results}
Figure~\ref{f:comp_workloads}(a) reports the latency of a single run, whereas Figure~\ref {f:comp_workloads}~(b) illustrates the amortized time across slots for FHE methods supporting SIMD parallelism (Scheme Switching and Encoding Switching).

TFHE~\cite{chillotti2020tfhe} evaluates all operations bit-wise, handling non-linear functions efficiently but performing poorly on linear ones at larger bit lengths. It lacks SIMD support, unlike word-wise schemes that process multiple ciphertexts simultaneously. Notably, TFHE performs better on workload-2 than workload-1 because, in workload-2, one multiplication input is the comparison result (all 0s or all 1s), reducing circuit depth compared to workload-1, where multiplication precedes comparison

Scheme Switching~\cite{boura2020chimera}, on the other hand, evaluates linear operations using word-wise FHE and non-linear operations using bit-wise FHE. However, the switching process's complexity grows exponentially with the input bit-width. For instance, on workload-1, 6-bit inputs take 9.87 seconds and 8-bit inputs 32.1 seconds, with runtimes for 12- and 16-bit inputs becoming prohibitive, making the approach impractical for larger inputs.

Encoding Switching~\cite{zhang2023hebridge} avoids the overhead of expensive scheme switching and leverages SIMD capabilities by performing both linear and non-linear operations in word-wise FHE. For 8-bit inputs, workload-1 takes only 15.5 seconds, and the method scales efficiently to 12- and 16-bit inputs (23.0 and 46.9 seconds), achieving up to 14.1$\times$ and 25.4$\times$ speedups over TFHE and scheme switching, respectively.

When comparing the amortized time in Figure~\ref{f:comp_workloads}~(b) with the single-run results, TFHE overtakes scheme switching as the slowest method, owing to its lack of SIMD support.

{\color{black}Table~\ref{tab:storage-comm} reports storage and communication metrics for the three general FHE methods on Workload-1 ($1\times100$ vector of 8-bit integers). TFHE achieves the smallest footprint (4.8 MB ciphertext, 9.6 MB communication, 23.3 MB memory), making it ideal for bandwidth-constrained scenarios, while Encoding Switching exhibits 10× larger overhead (47.6 MB, 95.2 MB, 94.3 MB) but enables faster computation in resource-rich environments. These metrics introduce deployment constraints into method selection: applications with limited bandwidth or storage should prioritize TFHE despite higher computational costs, whereas datacenter deployments can accept Encoding Switching's larger footprint for performance gains.

\begin{table}[h]
\centering
\small
\caption{Storage and communication overhead (Workload-1).}
\label{tab:storage-comm}
\begin{tabular}{@{}lccc@{}}
\toprule
\textbf{Method} & \textbf{Ciphertext} & \textbf{Peak Memory} & \textbf{Communication}$^{\ast}$ \\
\midrule
TFHE & 4.8 MB  & 23.3 MB & 9.6 MB  \\
Scheme Switching  & 14.2 MB & 58.6 MB & 28.4 MB \\
Encoding Switching& 47.6 MB & 94.3 MB & 95.2 MB \\
\bottomrule
\multicolumn{4}{l}{\footnotesize $^{\ast}$Communication between server and client.}
\end{tabular}
\end{table}
}



\section{General Computation Required Applications}
\label{sec:applications}

\subsection{Design Principle}
In this section, we evaluate general FHE methods on five representative privacy-sensitive AI applications that require general computation: private graph algorithms, private decision tree inference, private database aggregation, private neural network inference, and private sorting. These applications cover a diverse range of real-world use cases across domains such as healthcare, finance, and other industrial sectors. Each application incorporates both linear and non-linear operations and is built on the workloads analyzed in Section~\ref{sec:workloads}.


\subsection{Experimental Setup}
Generally, we evaluate various applications using Scheme Switching, Encoding Switching, and bit-wise TFHE, with experimental setup details identical to those described in Section~\ref{sec:workload_setup}. For private graph evaluation, a graph with \(n\) nodes is represented as an \(n\times n\) adjacency matrix. In the case of bit-wise TFHE, each element of the matrix is individually encrypted into multiple ciphertexts (for multiple bits), whereas for SIMD-enabled methods, each row is batched into a single ciphertext. For private decision tree evaluation, each node is encrypted as multiple ciphertexts for bit-wise FHE, while for SIMD-enabled methods, an entire layer is batched into one ciphertext. In private sorting, each element is encrypted into a single ciphertext as described in~\cite{iliashenko2021faster}. For private database aggregation, every element is encrypted into multiple ciphertexts when using bit-wise TFHE, whereas for SIMD-enabled methods, each column is batched into a single ciphertext. Finally, for private neural network inference, each parameter is encrypted into multiple ciphertexts with bit-wise TFHE, while for SIMD-enabled methods, each filter or weight matrix is batched into a single ciphertext.



\subsection{Private Graph}

Private graph algorithms~\cite{xie2014cryptgraph, dockendorf2022graph, meng2015grecs, li2021checking} enable secure computations on graph-structured data while preserving individual privacy~\cite{xie2014cryptgraph, mittal2012preserving, akcora2012privacy}. In this threat model, an untrusted server performs computations without learning any details about the input graph. A common operation in graph algorithms is selecting the shorter path from paths \(P_1\) and \(P_2\), which can be implemented as: \((P_1 > P_2) \times P_2 + (1 - (P_1 > P_2)) \times P_1\); this computation exemplifies workload-2, where a linear operation follows a non-linear operation. In this section, we demonstrate the performance of general FHE methods on the private Floyd-Warshall algorithm~\cite{floyd1962algorithm}.

\noindent\textbf{Floyd-Warshall on Plaintext.} The Floyd-Warshall algorithm is a classic algorithm for finding the shortest paths between all pairs of nodes in a graph \(G\). It returns on a distance matrix \(D\), where \(D[i,j]\) represents the shortest known distance from node \(i\) to node \(j\).

As shown in Algorithm~\ref{a:floyd-warshall-p}, the algorithm iteratively updates \(D\) by considering each node as an intermediate node in potential paths. During each iteration, it updates the distance \(D[i,j]\) for each pair of nodes \((i,j)\) by comparing it with \(D[i,k] + D[k,j]\), which represents a path from \(i\) to \(j\) via \(k\). This process is repeated for all nodes \(k\), ensuring all potential intermediate nodes are considered. After all iterations, the \(D\) contains the shortest paths between all pairs of nodes in the graph. The algorithm has a time complexity of \(O(|V|^3)\), where \(|V|\) is the nodes number.

\begin{algorithm}
\footnotesize
\caption{Floyd-Warshall Algorithm on Plaintext}
\begin{algorithmic}[1]
\STATE \textbf{Input:} $G$: adjacency matrix
\STATE \textbf{Output:} $D$: matrix of shortest path distances between each pair of nodes, $P$: matrix of predecessors for each node

\STATE \textit{\textcolor{black}{\# Initialize distance array $D$ and predecessor array $P$}}
\FOR{each vertex $v$ in $G$}
    \STATE $D[v] = G[v]$, $P[v] = \text{null}$
\ENDFOR

\STATE \textit{\textcolor{black}{\# Iteratively update the distances}}
\FOR{each vertex $k$ in $G$}
    \FOR{each vertex $i$ in $G$}
        \FOR{each vertex $j$ in $G$}
            \IF{$D[i, k] + D[k, j] < D[i, j]$}
                \STATE $D[i, j] = D[i, k] + D[k, j]$
                \STATE $P[i, j] = P[k, j]$
            \ENDIF
        \ENDFOR
    \ENDFOR
\ENDFOR

\STATE \textbf{return} $D$, $P$
\end{algorithmic}
\label{a:floyd-warshall-p}
\end{algorithm}

\noindent\textbf{Bit-wise FHE Private Floyd-Warshall.} To implement a private Floyd-Warshall algorithm, a naive method~\cite{dockendorf2022graph} encrypts each element of the adjacency matrix \(G\) to ciphertext. The server then conducts the same procedure on the encrypted \(G\) as delineated in Algorithm~\ref{a:floyd-warshall-p}. For if-else operations based on comparison results, the server uses HE multiplication to implement the conditional logic. Specifically, lines 10-11 of Algorithm~\ref{a:floyd-warshall-p} can be computed in HE as:
\begin{equation}
D[i,j] = M \cdot (D[i,k]+D[k,j]) + (1-M) \cdot D[i,j],
\end{equation}
where \(M\) is the encrypted comparison result. This naive private Floyd-Warshall requires \(|V|^3\) private comparisons and \(4|V|^3\) private multiplications, resulting in significant computational overhead.

\noindent\textbf{Word-wise FHE Private Floyd-Warshall with SIMD.} The cubic complexity of bit-wise naive private Floyd-Warshall stems from the three nested loops. However, we can optimize this process by leveraging the SIMD mechanism of word-wise FHE with the algorithm's inherent parallelism.

After fixing the middle node \(k\), updates for paths from node \(i\) to other nodes via \(k\) can be computed simultaneously. This parallelism allows SIMD by encoding the adjacency matrix row by row.


\begin{figure}[h!]
    \centering
    \includegraphics[width=\linewidth]{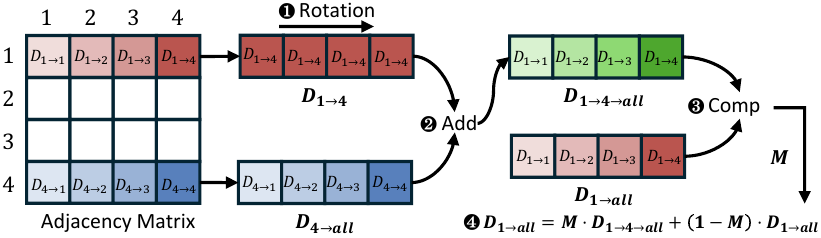}
    \caption{Toy example of SIMD update distances between node 1 to others through node 4.}
    \label{f:graph_multiple}
\end{figure}

Figure~\ref{f:graph_multiple} illustrates this SIMD optimization for a graph with \(4\) nodes, focusing on updating distances from node \(1\) to all others through node \(4\). The process is:
\ding{182} Generate ciphertext \(\boldsymbol{D_{1\rightarrow 4}}\) by multiplying \([D_{1\rightarrow 1}, D_{1\rightarrow 2}, D_{1\rightarrow 3}, D_{1\rightarrow 4}]\) with plaintext \([0,0,0,1]\), then replicating \(D_{1\rightarrow 4}\) across all slots by rotations and addition.
\ding{183} Compute \(\boldsymbol{D_{1\rightarrow 4\rightarrow all}}\) by adding \(\boldsymbol{D_{1\rightarrow 4}}\) to \(\boldsymbol{D_{4\rightarrow all}}\), representing new potential distances via node \(4\).
\ding{184} Compare \(\boldsymbol{D_{1\rightarrow 4\rightarrow all}}\) and \(\boldsymbol{D_{1\rightarrow all}}\), yielding mask \(\boldsymbol{M}\) (\(1\) where new distance is shorter, \(0\) otherwise).
\ding{185} Update distances by:
\begin{equation}
    \boldsymbol{D_{1\rightarrow all}}=\boldsymbol{M}\cdot \boldsymbol{D_{1\rightarrow4\rightarrow all}}+(1-\boldsymbol{M})\cdot \boldsymbol{D_{1\rightarrow all}},
\end{equation}
Only paths via node \(4\) that are shorter than the original one will be updated.


This SIMD-enabled approach, as shown in Algorithm~\ref{a:floyd-warshall-c}, allows each row of the adjacency matrix to be encrypted as a single ciphertext. Consequently, paths from node \(i\) to all other nodes via node \(k\) can be compared in a single operation, reducing the total number of HE comparisons from \(O(|V|^3)\) to \(O(|V|^2)\), and HE multiplications from \(O(4|V|^3)\) to \(O(4|V|^2)\).

\begin{algorithm}
\footnotesize
\caption{Floyd-Warshall Algorithm on Ciphertext Using SIMD}
\begin{algorithmic}[1]
\STATE \textbf{Input:} $G$: adjacency matrix (encrypted row-by-row)
\STATE \textbf{Output:} $D$: matrix of shortest path distances between each pair of nodes (encrypted), $P$: matrix of predecessors for each node (encrypted)

\STATE \textit{\textcolor{black}{\# Initialize distance array $D$ and predecessor array $P$}}
\FOR{each vertex $v$ in $G$}
    \STATE $D[v] = G[v]$, $P[v] = \text{null}$
\ENDFOR

\STATE \textit{\textcolor{black}{\# Iteratively update the distances using SIMD}}
\FOR{each vertex $k$ in $G$}
    \FOR{each vertex $i$ in $G$}
        \STATE $D_{i\rightarrow k} = \text{SIMD}(D[i,k])$ \textit{\textcolor{black}{\# Broadcast $D[i,k]$ to all slots}}
        \STATE $D_{k\rightarrow \text{all}} = D[k, :]$
        \STATE $D_\text{new} = D_{i\rightarrow k} + D_{k\rightarrow \text{all}}$
        \STATE $M =D_\text{new} < D[i,:]$ \textit{\textcolor{black}{\# Mask for minimum}}
        \STATE $D[i,:] = M \cdot D_{new} + (1 - M) \cdot D[i,:]$
        \STATE $P[i,:] = M \cdot P[k,:] + (1 - M) \cdot P[i,:]$
    \ENDFOR
\ENDFOR

\STATE \textbf{return} $D$, $P$
\end{algorithmic}
\label{a:floyd-warshall-c}
\end{algorithm}

\begin{figure}[h!]
    \centering
    \includegraphics[width=1\linewidth]{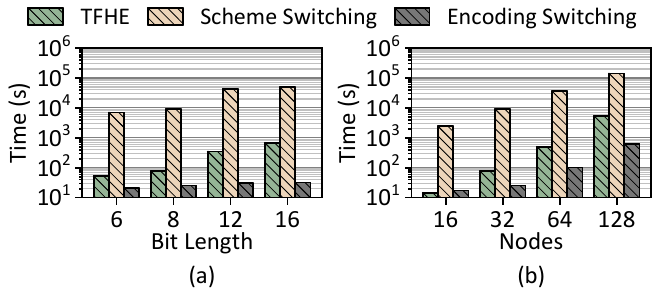}
    \caption{Homomorphic Floyd-Warshall. (a) Execution time for a 32-node graph with varying input bit lengths. (b) Execution time for graphs of different sizes with 8-bit input.}
    \label{f:results_floyd}
    \vspace{-0.25in}
\end{figure}

\noindent\textbf{Experimental Analysis.} 
To assess the performance of different general FHE methods for the Private Floyd-Warshall algorithm, we conducted experiments across various input bit widths and graph sizes. Figure~\ref{f:results_floyd}~(a) compares the performance for a graph with 32 nodes under different input bit lengths, where Encoding Switching exhibits the best performance and Scheme Switching the worst, particularly for larger bit widths. Figure~\ref{f:results_floyd}~(b) shows results for graphs of varying sizes using an 8-bit input. As noted earlier, the runtime of bit-wise TFHE grows approximately as \(O(|V|^3)\) due to the lack of SIMD support, while the complexities for both Scheme Switching and Encoding Switching are reduced to roughly \(O(|V|^2)\). For 8-bit inputs, TFHE takes 14.69 seconds, and Encoding Switching takes 17.28 seconds on a 16-node graph; however, for a 128-node graph, TFHE's runtime escalates dramatically to 5,432 seconds, whereas Encoding Switching completes in 632 seconds.




\subsection{Private Decision Tree Evaluation}

Private Decision Tree Evaluation (PDTE)~\cite{akavia2022privacy, kiss2019sok, akhavan2023level, azogagh2022probonite} allows a server to provide predictions using a private decision tree on a client's confidential input, preserving both input privacy (the server learns nothing about the client's data) and model privacy (the client only obtains the inference result without learning any details of the decision tree). 

Decision trees primarily involve comparisons and traversals~\cite{song2015decision}. In these trees, comparisons are made between the client's inputs and decision thresholds, yielding ciphertext values \(c_i\in[0,1]\) at each node, while traversals determine the active path by computing the product \(c_1\cdot c_2\cdot \dots \cdot c_d\), where \(c_i\) is the comparison result at the \(i\)-th level and \(d\) is the tree's depth. The correct decision is identified by the leaf node that decrypts to 1, with all other leaves decrypting to 0. {\color{black}Since private decision tree evaluation involves both linear and non-linear operations, it requires general FHE methods. Specifically, linear operations following non-linear comparisons correspond to workload-2 defined in Section~\ref{sec:workloads}.} Notably, the traversal operation \(c_1\cdot c_2\cdot \dots \cdot c_d\) can be reformulated~\cite{akhavan2023level} as:
\begin{equation}
c_1\cdot c_2\cdot \dots \cdot c_d = \overline{\overline{c_1} + \overline{c_2} + \dots + \overline{c_d}}
\label{e:convert}
\end{equation}
which significantly reduces the required multiplicative depth, further improving the efficiency, especially for decision trees with larger depths.

\begin{figure}[h!]
    \centering
    \includegraphics[width=1\linewidth]{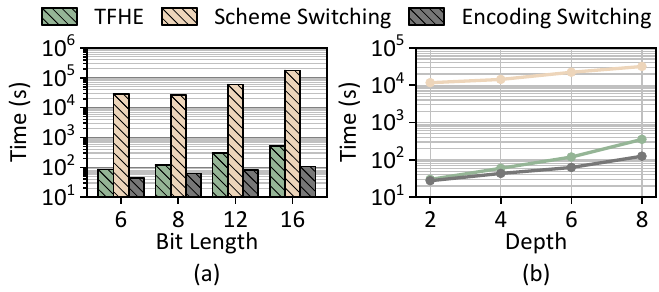}
    \caption{Private Decision Tree. (a) Execution time of inference for a depth-6 private decision tree under varying input bit precision; (b) Execution time of private decision trees with different depths using 8-bit input.}
    \label{f:results_decision_tree}
\end{figure}

\noindent\textbf{Experimental Analysis.}
In Figure~\ref{f:results_decision_tree}~(a), we evaluate the performance of various general FHE methods on a complete decision tree with a depth of 6 across different input bit lengths. Figure~\ref{f:results_decision_tree}~(b) shows the performance for complete decision trees of varying depths using an 8-bit input. The results indicate that for deeper trees, the word-wise Encoding Switching method performs more efficiently than others since it enables SIMD so that it can compare the same input with multiple thresholds simultaneously while avoiding the high cost of scheme switching between bit-wise and word-wise ciphertexts. On the other hand, the word-wise Scheme Switching performs worst because of the significant frequency of switching, since every non-linear operation follows a linear operation.

\subsection{Private Sorting}
\label{s:private_sort}

Private sorting is a protocol that enables secure ordering of encrypted data without revealing the underlying values or the sorting process. In this scenario, a client encrypts its data and sends it to a server, which then performs the sorting directly on the ciphertexts. Private sorting is critical and commonly used in privacy-sensitive areas such as financial. \textcolor{black}{For instance, several financial institutions might want to merge and sort transaction data to detect market trends without exposing individual client details.}

The most efficient homomorphic sorting algorithm in terms of running time is the direct sorting algorithm proposed by {\c{C}}etin, et al.~\cite{ccetin2015depth}. This approach directly determines the final position of each element by counting the number of elements less than it. Specifically, for an array of size \(m\), the server performs \(m(m-1)/2\) homomorphic less-than operations and an equal number of homomorphic additions on the less-than operations' result for computing the Hamming weight, which yields the correct indices \(I\) for each element.

Once the indices are established, the elements are repositioned in one efficient step using a conditional multiplexer (CMUX)-like operation, as follows:

\begin{equation}
S[i] = \sum_{j=1}^{m} X[j] \cdot \text{EQ}(i, I[j]),
\label{e:direct_sort}
\end{equation}
where \(S[i]\) is the element at the position \(I\) of sorted array \(S\) and \(X\) is the input array. \(\text{EQ}(i, I[j])\) is 1 if the \(i\) equals the index \(I[i]\) corresponding to \(X[j]\), and 0 otherwise. Constructing one position \(S[i]\) requires \(m\) equality evaluation and \(m\) multiplication, so sorting the entire encrypted array requires \(O(m^2)\) non-linear operations (less-than and equality evaluation) and \(O(m^2)\) linear operations (additions and multiplications).

{\color{black}Thus, homomorphically sorting an encrypted array of \(m\) elements demands general FHE methods, particularly as the linear operations are performed on the results of non-linear operations, corresponding to our workload-2.}

\begin{figure}[h!]
    \centering
    \includegraphics[width=\linewidth]{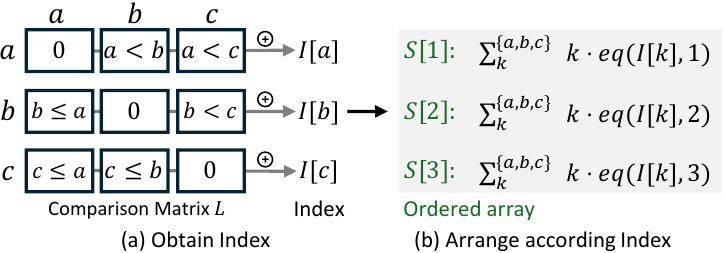}
    \caption{Toy example of direct sorting on the array \([a, b, c]\). (a) demonstrates the computation of the Comparison Matrix \(L\) and the derivation of indices from the hamming weight of \(L\), where each element's index corresponds to the count of elements greater than it. (b) shows the sorting mechanism where each element is multiplied by a binary value: only the element with a matching index retains its value (multiplied by 1), while others are set to zero (multiplied by 0).}
    \label{f:direct_sort}
\end{figure}

Figure~\ref{f:direct_sort} shows an example to private sort a three-dim array \(X=[a,b,c]\), the server first computes a less-than matrix \(L\) by:
\begin{equation}
    \mathbf{L}_{ij} =
    \begin{cases}
        \text{LT}(x_i, x_j) & \text{if } i < j, \\
        0 & \text{if } i = j, \\
        1 - \text{LT}(x_j, x_i) & \text{if } i > j.
    \end{cases}
\end{equation}

Next, the server computes the Hamming weight of each row, i.e., \(i\)-th row, to obtain how many elements are larger than \(x_i\). The Hamming weight, i.e., \(I[a]\), \(I[b]\) and \(I[c]\), denote the indices of \(a\), \(b\) and \(c\) in the sorted array, respectively.
Finally, the sorted array \(S\) is constructed by selecting elements \(X[j]\) where the index matches the hamming weight, with Equation~\ref{e:direct_sort}.

\begin{figure}[h!]
    \centering
    \includegraphics[width=\linewidth]{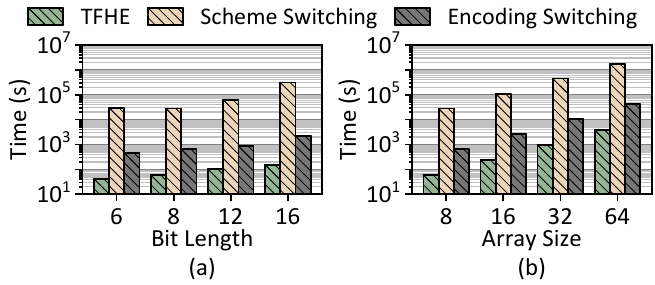}
    \caption{Homomorphic Sorting. (a) Execution time of sorting an 8-element array under varying input bit precision; (b) Execution time of sorting arrays of different lengths with 8-bit input.}
    \label{f:results_bubble_direct}
\end{figure}

\noindent\textbf{Experimental Analysis.} We conduct experiments using various general FHE methods on the private sorting tasks. Figure~\ref{f:results_bubble_direct}~(a) presents the performance of different FHE methods under different input bit precisions, and Figure~\ref{f:results_bubble_direct}~(b) shows the performance under different array dimensions. For instance, sorting an 8-element array requires 59.9 seconds with TFHE, compared to 645 seconds for Encoding Switching and 27,457 seconds for Scheme Switching. This efficiency advantage is due to two key factors. First, unlike other applications, private direct sorting on a single array cannot leverage SIMD~\cite{iliashenko2021faster}, which diminishes the benefits of word-wise methods like Scheme Switching and Encoding Switching. Second, in this context, one of the operands in the linear operations is the output of a non-linear operation (either 0 or 1), resulting in a simpler circuit compared to linear operations on arbitrary integers. Conversely, Scheme Switching shows the lowest efficiency because its frequent switching, triggered by each non-linear operation following a linear one, adds significant overhead.

The reasons are two-fold; firstly, unlike other applications, private direct sorting of an encrypted array cannot leverage the SIMD~\cite{iliashenko2021faster}, which disables the advantages of word-wise Scheme Switching and Encoding Switching. Secondly, one of the linear operation operands is the result of a non-linear operation, i.e., 0 or 1. So that the TFHE circuit would be much less than linear operation between two random integers. On the other hand, Scheme Switching presents the lowest efficiency across all input bits and array sizes because of the frequent switching, since every non-linear operation will follow a linear operation.

\subsection{Private Database Aggregation}
\label{s:private_database}
Private Database Aggregation (PDBA) is a protocol that protects client privacy during aggregate queries on cloud-stored databases, such as Microsoft Azure SQL Server~\cite{mircroSQL} and AWS Aurora~\cite{AWSAurora}. Recently, Leidos has partnered with AWS to explore the solution of using FHE to protect databases~\cite {fhe_database}. It allows clients to securely store encrypted data on a cloud server and perform operations, like summation, without disclosing data or the details of the queries. The server processes these functions on encrypted data and returns only the encrypted aggregate results, ensuring the server does not access actual data or query specifics.

Recently, researchers have developed encrypted databases to ensure data privacy while enabling encrypted data processing. Cryptography-based solutions~\cite{popa2011cryptdb, tu2013processing, papadimitriou2016big, hackenjos2020sagma}, utilize cryptographic primitives such as DET~\cite{bellare2007deterministic}, SE~\cite{cash2013highly,curtmola2006searchable}, partially PHE~\cite{paillier1999public, rivest1978method} and order OPE/ORE~\cite{chenette2016practical, boldyreva2009order}. Recent advancements have explored the application of FHE in HEDA~\cite{ren2022heda}, HE3DB~\cite{bian2023he3db}, ArcEDB~\cite{zhang2024arcedb}, and Engorgio~\cite{bian2025engorgio}.

An aggregation query typically involves a combination of linear operations and non-linear operations. Consider the following SQL query as an example:

\begin{lstlisting}
SELECT ID FROM emp
    WHERE salary * work_hours BETWEEN 5000 AND 6000
          AND salary + bonus BETWEEN 700 AND 800;
\end{lstlisting}

{\color{black}In this query, the filtering predicates "salary * work\_hours BETWEEN 5000 AND 6000" and "salary + bonus BETWEEN 700 AND 800" involve a non-linear operation following a linear operation. This requirement necessitates general FHE methods and aligns with workload-1 defined in Section~\ref{sec:workloads}.}

While word-wise FHE methods work well for linear operations, they are less effective for non-linear operations like comparisons and logic filtering. To address this limitation and leverage the high efficiency of SIMD, Ren et al. introduce a novel framework, HEDA~\cite{ren2022heda}. The core idea is to utilize scheme switching between RLWE and LWE ciphertexts. This approach conducts non-linear operations on LWE ciphertexts, where such operations are more naturally supported. Then, it converts the results back to RLWE ciphertexts to perform linear operations in a SIMD fashion.

\begin{figure}[h!]
    \centering
    \includegraphics[width=\linewidth]{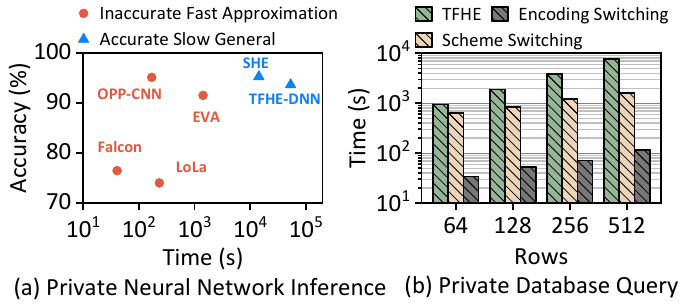}          
    \caption{(a) compares the performance of various FHE DNN works on CIFAR-10 using VGG-9. (b) Private database implemented by different schemes.}
    \label{f:results_nn}
\end{figure}

\noindent\textbf{Experimental Analysis.} As described in Section~\ref{s:private_database}, SQL queries involve combinations of non-linear and linear operations. To evaluate the performance of each general FHE method on the private database query, we conducted experiments on Query~\ref{lst:sql_query} across databases with different numbers of rows under 8-bit precision, as illustrated in Figure~\ref{f:results_nn}~(b).

The results show that the word-wise general FHE methods generally outperform the bit-wise TFHE, largely due to their use of SIMD capabilities, a feature that the bit-wise TFHE lacks. Furthermore, as the number of rows increases, Encoding Switching demonstrates better efficiency than Scheme Switching. For example, Encoding Switching takes 115 seconds for a 512-row database, compared to 1,626 seconds for Scheme Switching. This advantage is attributed to Encoding Switching's ability to perform non-linear operations in a SIMD-style, while Scheme Switching processes non-linear operations on bit-wise ciphertexts and cannot leverage SIMD.

\subsection{Private Neural Network}
\label{s:private_nn}
One of the most valuable applications of FHE is in facilitating privacy-preserving deep learning, i.e., Deep Learning as a Service (DLaaS)~\cite{srinivasan2019delphi,mohassel2017secureml}, which is commonly used in information-sensitive areas such as financial~\cite {han2019logistic} and healthcare~\cite{kim2015private}. This scenario involves a server that possesses a deep learning model, i.e., a neural network, and a client who encrypts their input data before sharing it with the untrusted server. The server processes encrypted input to provide predictions without accessing any sensitive information, thereby ensuring a secure inference service.

{\color{black}Neural networks consist of linear layers such as convolutions and non-linear layers like ReLU and MaxPool~\cite{krizhevsky2012imagenet, lecun1998gradient}. The combination of linear and non-linear operations required by these layers necessitates general FHE methods; indeed, the iterative sequence of linear and non-linear layers forms the basis of workload-3 as defined in Section~\ref{sec:workloads}.}

Despite the use of general FHE methods, some research leveraged the specific feature of neural networks to improve efficiency, but sacrificed some accuracy. Specifically, polynomial approximation-based methods~\cite{lee2022privacy, lee2023precise, lee2022optimization} leverage the noise resilience of neural networks to enhance efficiency while reducing multiplicative depth and complexity. Although these approximations are typically accurate only within a narrow range (often between -1 and 1), activation layer inputs are usually constrained (e.g., within \([-32, 32]\) or \([-64, 64]\)). By scaling these inputs to fit within \([-1, 1]\) using a predetermined scalar, servers can effectively implement these approximation techniques throughout the network, as demonstrated in approaches like DaCapo~\cite{cheondacapo}, OPT-CNN~\cite{kim2023optimized}, and Hetal~\cite{lee2023hetal}, among others~\cite{lee2022low, dathathri2020eva}. However, determining the appropriate scaling factor is critical: if it is too large, inputs may not be sufficiently constrained within \([-1, 1]\); if too small, the inputs become overly diminished, resulting in significant approximation errors near 0.

Another approximation strategy simulates non-linear activation functions using linear operations. For example, CryptoNets~\cite{gilad2016cryptonets} replaces ReLU with the square function, a method later adopted by frameworks like LoLa~\cite{brutzkus2019low} and Falcon~\cite{lou2020falcon}. This approach avoids the overhead of homomorphic non-linear operations by training the neural network with a substitute activation function.

\noindent\textbf{Experimental Analysis.} To evaluate the performance of various private neural network inference implementations described in Section~\ref{s:private_nn}, we conducted experiments using the CIFAR-10 dataset~\cite{krizhevsky2009learning} on a VGG-9 model. VGG-9 is a streamlined version of the VGG network family, comprising 9 layers including convolutional and fully connected layers. Typically, it consists of 6 convolutional layers followed by 3 fully connected layers, with ReLU activations and max pooling operations interleaved between the convolutional layers. Encryption parameters were set as specified in each referenced paper to ensure at least 128-bit security.

Figure~\ref{f:results_nn}~(a) shows that implementations using the square function as an activation substitute, such as Falcon~\cite{lou2020falcon} and LoLa~\cite{brutzkus2019low}, yield lower performance. This is primarily because training with degree-2 polynomial activations can lead to instability issues like exploding or vanishing gradients~\cite{ali2020polynomial,santriaji2024dataseal,  xue2022audit, xue2025dictpflefficientprivatefederated, zheng2023cryptography}. Although the square function enables faster inference, it is generally only suitable for smaller, less complex neural networks.

In contrast, private neural networks that employ approximation-based activation functions, such as OPP-CNN~\cite{kim2023optimized} and EVA~\cite{dathathri2020eva}, achieve higher accuracy levels. Meanwhile, TFHE's performance is significantly slower compared to implementations using interpolation approximations. This discrepancy arises because neural network architectures predominantly consist of linear layers, which benefit from the SIMD capabilities of word-wise CKKS schemes. Bit-wise TFHE, lacking SIMD support and optimized linear operations, exhibits lower performance. Additionally, the Scheme Switching method (SHE~\cite{lou2019she}) achieves the highest accuracy but at slower inference times than the approximation-based approaches.


\section{Recommendation and Discussion}

\begin{figure}[t!]
    \centering
    \includegraphics[width=\linewidth]{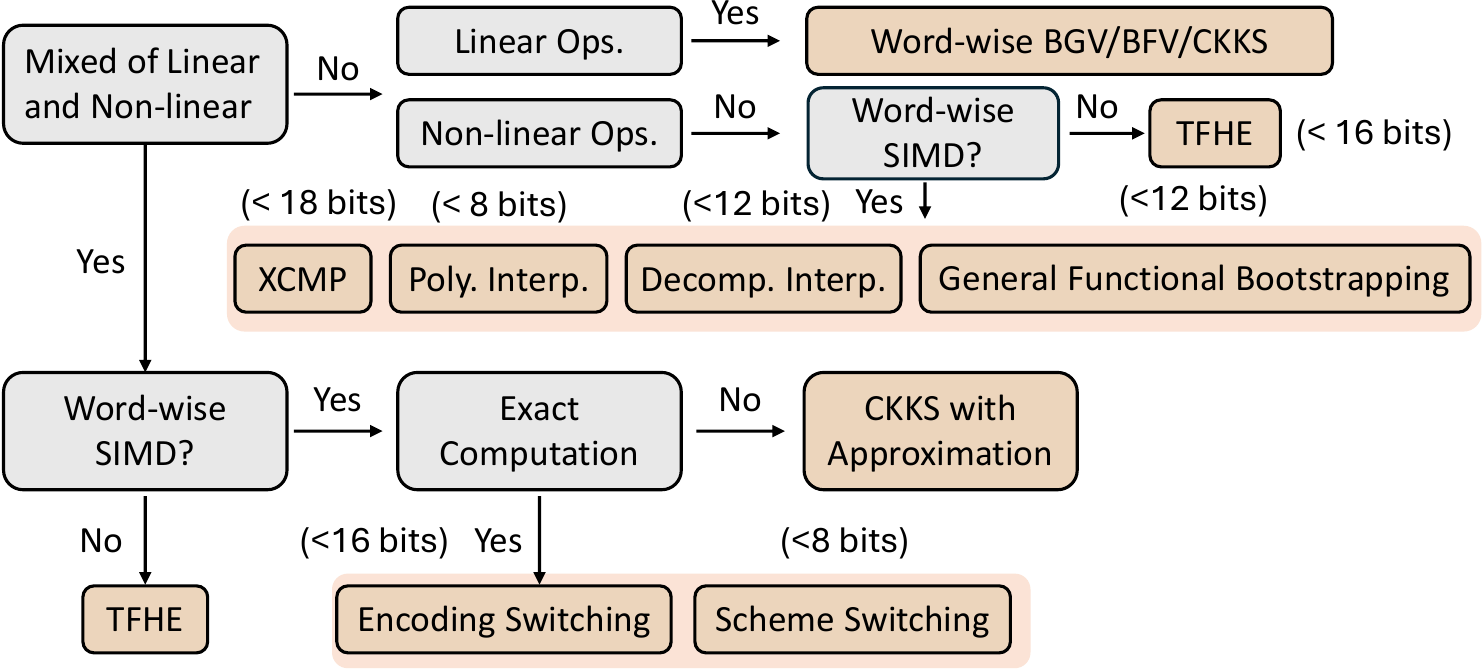}
    \caption{Flow chart to guide users to the FHE method that best fits the requirements of their privacy-sensitive applications and available resources.}
    \label{f:flowchart}
\end{figure}

Considering the capabilities of each FHE method, we offer recommendations tailored to the types of operations required by a given privacy-preserving application and whether parallelism can be exploited. Figure~\ref{f:flowchart} summarizes these guidelines.

First, if the application involves only linear operations, word-wise FHE methods (BGV, BFV, CKKS) are the most efficient choice. Conversely, if the application requires only non-linear operations, the next consideration should be whether the application can benefit from the word-wise SIMD technique. If SIMD does not offer an advantage, a bit-wise FHE method, such as TFHE, is the most efficient option. However, if SIMD is useful, several techniques enable non-linear functionalities in a word-wise context, including direct polynomial interpolation~\cite{narumanchi2017performance}, decomposition-based polynomial interpolation~\cite{iliashenko2021faster}, general functional bootstrapping~\cite{geelen2023bootstrapping}, or special encoding methods like XCMP~\cite{akhavan2023level}. These word-wise methods enable parallel computation to improve efficiency.

When an application demands both linear and non-linear computations (i.e., computational generality-required applications), the next question is whether the workload can benefit from SIMD parallelism. If SIMD does not offer advantages, e.g., private sorting for an array, a bit-wise FHE method, such as TFHE, remains the most practical choice. If SIMD is beneficial, the final consideration is whether exact, error-free results are necessary. If approximate results are acceptable, as is common in tasks like neural network inference, where noise tolerance is inherent, CKKS with approximation~\cite{cheon2020efficient, lee2021minimax} offers the most efficiency. If exact computation is required, the Encoding Switching~\cite{zhang2023hebridge} between beFV and FV, and the Scheme Switching~\cite{lu2021pegasus} between word-wise and bit-wise FHE can deliver the best performance.

{\color{black}
While our SoK focuses on pure FHE methods for non-interactive general computation, we acknowledge related approaches with different trade-offs. Hybrid FHE-MPC protocols~\cite{xue2023cryptotrain, huang2022cheetah, he2024rhombus, hao2022iron, pang2024bolt} improve efficiency by using FHE for linear operations and MPC for non-linear operations. However, these hybrid approaches require frequent client-server communication during computation, sacrificing non-interactivity. In contrast, the pure FHE methods we focus on enable servers to complete the entire computation cycle independently, without requiring the client to remain online. This non-interactive property is critical for bandwidth-constrained environments.

Another research direction is optimizing rotation key management to reduce client-to-server transmission~\cite{cheon2025towards, lee2023rotation} or accelerate linear operations~\cite{zheng2023new, aikata2024secure} via specical encoding. While valuable, these works are orthogonal to our focus: we systematically analyze which FHE methods achieve functional generality supporting arbitrary combinations of linear and non-linear operations non-interactively, thereby establishing the fundamental computational capabilities and limitations of pure FHE for general AI workloads.
}


\section{Conclusion}
In this paper, we systematically evaluate the computational generality of Fully Homomorphic Encryption (FHE)—its ability to perform non-interactive mixed linear and non-linear operations—a critical requirement for real-world privacy-sensitive applications like neural networks, graph analytics, and secure database queries. Through a three-stage methodology combining theoretical complexity analysis, micro-benchmarks on mixed-operation workloads, and empirical evaluations across five applications, we reveal significant overheads in existing general FHE solutions, emphasizing the urgent need for optimizations that bridge efficiency gaps while preserving security. To guide practitioners, we provide recommendations for selecting optimal FHE methods based on operation types, parallelism opportunities, and error tolerance, enabling developers to balance efficiency, precision, and security in privacy-preserving systems.

\begin{acks}
This work was partially supported by NSF CSR-2413232, NSF SaTC-2523407, and student internship support from Samsung Research America. Any opinions, findings and conclusions or recommendations expressed in this material are those of the authors and do not necessarily reflect the views of grant agencies or their contractors.
\end{acks}

\bibliographystyle{plain}
\bibliography{main}
\appendix
\section{Appendix}

\subsection{CKKS with Polynomial Approximation}
\label{app:CKKS}
CKKS is a word-wise FHE method capable of encrypting real numbers. Since it supports approximate computation on floating-point numbers, the most common method for performing non-linear functions is polynomial approximation~\cite{lee2023precise, lee2022privacy}. Specifically, these methods use compositions of minimax approximation polynomials to approximate the sign function over a small interval, typically \([-1, -\epsilon] \cup [\epsilon, 1]\), with errors. Given an $n$-degree approximation polynomial, the complexity is \(O(\sqrt{n})\) and the multiplicative depth is \(\log_2n\). The larger the interval and the smaller the error, the higher the polynomial degree required---thus increasing the number of multiplications and multiplicative depth. Most importantly, approximation can never be accurate around \(0\). These errors limit the applicability in domains where accuracy is critical, such as genomics~\cite{kim2015private, raisaro2018protecting, zhang2015foresee} and finance~\cite{armknecht2015guide, han2019logistic}.

Overall, polynomial approximation strikes a balance between computational efficiency and accuracy, making it well-suited for applications such as neural network inference, where approximate computations are acceptable.

\subsection{BGV/BFV with Polynomial Interpolation}
\label{app:Interpolation}
In contrast to CKKS, computations in BFV/BGV are exact. Accordingly, non-linear functions can be implemented in BFV/BGV without approximation error. Narumanchi et al.~\cite{narumanchi2017performance} proposed computing non-linear functions via polynomial interpolation.

Take the non-linear comparison function as an example, since the comparison is a fundamental operation for many logic functions and is key to evaluating various non-linear operations~\cite{zhang2023hebridge}. The core idea is to convert the comparison between two encrypted integers, \(a\) and \(b\), into a comparison between their difference, \( z = b - a \), and zero. This is achieved by constructing a polynomial \(P(x)\) with the property:

\begin{equation}
    P(x) =
    \begin{cases}
        1, & \text{if } x < 0;\\
        0, & \text{if } x \geq 0;
    \end{cases}
\label{e:p}
\end{equation}
\( P(x) \) are expressed using Lagrange interpolation:
\begin{equation}
   P(x) = \sum_{i=0}^p \left(\prod_{j=0, j \neq i}^p \frac{x - x_j}{x_i - x_j}\right) \cdot y_i \mod p,
\end{equation}
where \( x_i \) and \( y_i \) are chosen such that \( y_i = 1 \) for \( x_i < 0 \) and \( y_i = 0 \) for \( x_i \geq 0 \). Once the polynomial is constructed, it is evaluated homomorphically on the encrypted difference \( \text{Enc}(a-b) \):
\begin{equation}
    \text{Enc}(z) = P(\text{Enc}(a-b)).
\end{equation}

Upon decryption, \(\text{Dec}(\text{Enc}(z)) = 1\) indicates \(a<b\), while \(\text{Dec}(\text{Enc}(z)) = 0\) indicates \(a \geq b\).

The polynomial interpolation has notable limitations. Specifically, the standard BGV and BFV support efficient linear operations over \(\mathbb{Z}_{p^r}\)~\cite{chen2018homomorphic, halevi2021bootstrapping}. However, polynomial interpolation in \(\mathbb{Z}_{p^r}\) is impractical, as it leads to polynomials of degree \(O(p^r)\), which grow exponentially with the input bit-width. Evaluating such high-degree polynomials is prohibitive in practice~\cite{iliashenko2021faster, tan2020efficient}. Therefore, non-linear operations are typically performed in the base field \(\mathbb{Z}_p\). The degree of interpolation polynomials equals \(p\), requiring \(O(\sqrt{p})\) non-scalar multiplications and \(O(\log_2p)\) multiplicative depth using the Paterson-Stockmeyer algorithm. To enable efficient evaluation, the plaintext space is typically confined to a small prime \(p\), making it unsuitable for large inputs. The value of \(p\) ranges from \(2\sim 7\) in~\cite{tan2020efficient}, is at most \(173\) in~\cite{iliashenko2021faster} and at most \(257\) in~\cite{morimura2023accelerating}. In conclusion, this limitation makes it impractical for real-world AI applications which require general computation.


\subsection{Polynomial Interpolation with Special Encoding}
\label{app:SE}
Aiming to enhance the capability of polynomial interpolation to support non-linear operations on large inputs, several special encoding-based methods have been proposed~\cite{iliashenko2021faster, tan2020efficient}, which encode a large integer as a vector of base-$p$ digits: \(a = (a_0, a_1, \ldots, a_{r-1})\), where each \(a_i \in \mathbb{Z}_p\). Consequently, a non-linear operation, e.g., a comparison between two integers \(a\) and \(b\), is transformed into a lexicographical comparison of their respective digit vectors. Formally, the process begins with the decomposition of the input integers \(a\) and \(b\) into their digit representations:
\begin{equation}
\begin{aligned}
    \text{Enc}(a) &= (\text{Enc}(a_0), \text{Enc}(a_1), \ldots, \text{Enc}(a_{r-1})), \\
    \text{Enc}(b) &= (\text{Enc}(b_0), \text{Enc}(b_1), \ldots, \text{Enc}(b_{r-1})).
\end{aligned}
\end{equation}

For each digit pair \(a_i\) and \(b_i\), a polynomial \(P(\cdot)\) is constructed to evaluate the comparison \(a_i < b_i\) as Equation~\ref{e:p}. This polynomial is then applied homomorphically:
\begin{equation}
\text{Enc}(z_i) = P(\text{Enc}(a_i - b_i)).
\end{equation}

The overall comparison is obtained by combining these digit-wise comparisons lexicographically:

\begin{equation}
\text{Enc}(z) = \text{Enc}(z_0) + \sum_{i=1}^{r-1} \text{Enc}(z_i) \prod_{j=0}^{i-1} (1 - \text{Enc}(z_j)),
\label{e:decompose}
\end{equation}
where \(\prod_{j=0}^{i-1} (1 - \text{Enc}(z_j))\) ensures that less significant digits only contribute when more significant digits are equal.




By decomposing the large field into smaller subfields and performing comparisons at the digit level, this method achieves a significantly lower circuit depth. The depth of the circuit evaluating the comparison of two \(b\)-bit integers is given by:

\begin{equation}
\log_2\log_p2^b+\log_2(p-1)+4.
\end{equation}

In contrast, directly interpolation on \(\mathbb{Z}_{p^r}\) has a multiplicative depth of \(O(\log_2p^r)\). This reduced depth makes the method more practical for larger fields. By addressing the challenge of deep circuit depth in polynomial interpolation, \cite{iliashenko2021faster} enables faster homomorphic non-linear operations for practical applications in BGV/BFV. Despite its significant efficiency improvement, the special encoding-based ciphertexts do not support linear operations since the number is expressed in the vector form, making it not an FHE method enabling general computation.

Similarly, other special encoding-based approaches have also been proposed to enhance the efficiency of non-linear operations in word-wise FHE. Unfortunately, these methods sacrifice generality, as the specialized ciphertext formats do not support linear operations. Below, we present a representative example.

\subsection{Exponential Encoding (XCMP)}
\label{app:XCMP}
The exponential encoding method for FHE private comparison, named XCMP, was first proposed by Lu et al.~\cite{lu2018non}. The core idea is based on the fact that the message space in HE is a polynomial ring, i.e., \(\mathbb{Z}_p[x]/\langle x^n+1\rangle\), and that ciphertext multiplication corresponds to polynomial multiplication. This enables the design of a special encoding scheme that encodes the values to be compared into the polynomial's degree coefficients.

The comparison process of integers \(a\) and \(b\) begins with encoding \(a\) and \(-b\) as polynomials \(X^a\) and \(X^{-b}\), respectively. These polynomials are encrypted into ciphertexts \(\text{Enc}(X^a)\) and \(\text{Enc}(X^{-b})\), which are used to perform:
\begin{equation}
    \text{Enc}(C(X)) = T(X) \times \text{Enc}(X^a) \times \text{Enc}(X^{-b}) \mod (X^n + 1)
\label{e:xcmp}
\end{equation}
where \(T(X) = 1 + X + \cdots + X^{n-1}\). Note that the polynomial with a negative degree \(X^{-b} \equiv -X^{n-b} \mod (X^n + 1)\).

Finally, decrypt \(\text{Enc}(C(X))\) and evaluates the 0-th coefficient \(C(X)[0]\), which indicates the comparison result:

\begin{itemize}
    \item If \(a \leq b\), \(X^{b-a}\) from \(T(X)\) aligns with \(X^{a-b}\), yielding \(C(X)[0] = 1\).
    \item If \(a > b\), the \((n - (a - b))\)-th term of \(T(X)\) aligns with \(X^{a-b}\), resulting in \(C(X)[0] = -1\) due to the wrap-around: \(X^{n-(a-b)} \times X^{a-b} \equiv -1 \mod (X^n + 1)\).
\end{itemize}
In other words, \(C(X)[0] = 1\) if \(a \leq b\), and \(C(X)[0] = -1\) if \(a > b\).

XCMP has a multiplicative depth and complexity of 1 due to its use of a single ciphertext multiplication. However, it is limited by a small input field, requiring both \(a\) and \(b\) to be smaller than the polynomial degree \(n\). This constraint reduces efficiency for large inputs, as a larger \(n\) is needed~\cite{lu2018non}. Although there have been discussions about extending the domain to \(\mathbb{F}_{n^2}\) by decomposing numbers in base \(n\) and processing digits separately:

\begin{equation}
\mathbb{I}[a\leq b] = \mathbb{I}[a_1\leq b_1] + \mathbb{I}[a_1=b_1]\cdot \mathbb{I}[a_0\leq b_0],
\end{equation}
where the equality check result $\mathbb{I}[a_1=b_1]$ must be converted to the XCMP format for multiplication with $\mathbb{I}[a_0\leq b_0]$. This conversion involves Fermat's little theorem, resulting in a large multiplicative depth. As the result, XCMP performs well for small domains (less than 16 bits). For larger domains, performance significantly decreases~\cite{lu2018non}.

\subsection{General Functional Bootstrapping}
\label{app:GB}
Functional bootstrapping~\cite{alexandru2024general, lee2024functional, liu2023amortized} extends traditional bootstrapping techniques by not only refreshing ciphertexts but also by enabling the homomorphic evaluation of arbitrary functions on encrypted data. The key idea is to express a target function as a look-up table (LUT) and then approximate this LUT via a constructed interpolation polynomial. Lee et al.~\cite{lee2024functional} proposed functional bootstrapping for BFV, building upon regular BFV-style bootstrapping~\cite{geelen2023bootstrapping}. However, BFV-style bootstrapping efficiently supports only a small number of slots. To improve efficiency, Alexandru et al.~\cite{alexandru2024general} employed CKKS-style bootstrapping to increase slot utilization. Specifically, the target function is first encoded as a LUT over a finite base-\(p\) field \(\mathbb{Z}_p\), and then approximated using a trigonometric Hermite interpolation polynomial \(R(x)\). For instance, in the first-order case, the interpolation is constructed so that:
\begin{equation}
    R\left(\frac{k}{p}\right) = f(k) \:\: \text{for all} \:\: k \in \{0, 1, \dots, p-1\} \:\: \text{and} \:\: R'\left(\frac{k}{p}\right) = 0
\end{equation}

This polynomial is then evaluated homomorphically using an efficient technique such as the Paterson–Stockmeyer algorithm. The complexity of evaluating a polynomial of degree \( d \) using this method is roughly proportional to \( \sqrt{2d} + \log d \). For a first-order interpolation, where \( d = p-1 \), the complexity increases approximately as \( \sqrt{p} \). Moreover, if further noise reduction is required, higher-order interpolations can be used (e.g., second-order or third-order), which increase the polynomial degree (to about \( \tfrac{3p}{2} \) or \( 2p-1 \), respectively) and thus the overall cost. Moreover, extending the input space from \(\mathbb{Z}_p\) to \(\mathbb{Z}_{p^r}\) requires representing each integer as a vector of base‑\(p\) digits, similar to special encoding-based polynomial interpolation in BFV/BGV~\cite{zhang2023hebridge, iliashenko2021faster}, which precludes the continuous evaluation of linear and non‑linear functions on large inputs. Experiments in~\cite{alexandru2024general} demonstrate that general functional bootstrapping remains practical only for LUT evaluations up to 12‑bit inputs.

\end{document}